\documentclass[12pt]{article}

\renewcommand{\title}[1] {\begin{center}{\large\bf#1\bigskip}\end{center}}
\renewcommand{\author}[1] {\begin{center}#1\end{center}}
\newcommand{\address}[1] {\begin{center}{\sl #1\bigskip}\end{center}}
\newcommand{\lsim} {\buildrel < \over {_\sim}}
\newcommand{\gsim} {\buildrel > \over {_\sim}}
\newcommand{\VEV}[1]{\left\langle{#1}\right\rangle}

\begin{document}

\flushbottom \setcounter{footnote}{0}
\renewcommand{\thefootnote}{\fnsymbol{footnote}}

\begin{flushright}
UMD--02--047\\
SLAC--PUB--9204 \\
April 2002
\end{flushright}
\vfill

\title{CONFRONTING THE CONVENTIONAL IDEAS OF GRAND\\[.25ex] UNIFICATION WITH
FERMION MASSES, NEUTRINO \\[.6ex] OSCILLATIONS AND PROTON
DECAY\footnote{Invited talk presented at the International Summer
School held at ICTP, Trieste (June, 2001) and at WHEPP-7
Conference, Allahabad, India (January, 2002).  This is an updated
version of the talk presented at the Erice School (September,
2000), hep-ph/0106082}}

\vfill

\author{Jogesh C. Pati}
\address{Department of Physics,  University of Maryland,
College Park  MD 20742\footnote{Permanent address.}\\
and\\
Stanford Linear Accelerator Center, Menlo Park, CA 94025}

 \flushbottom \setcounter{footnote}{0}
\renewcommand{\thefootnote}{\arabic{footnote}}

\begin{abstract}

It is noted that one is now in possession of a set of facts, which
may be viewed as the {\em matching pieces of a puzzle}; in that
all of them can be resolved by just one idea---that is grand
unification. These include (i) the observed family-structure, (ii)
quantization of electric charge, (iii) the meeting of the three
gauge couplings, (iv) neutrino oscillations [in particular the
value $\Delta m^2(\nu_\mu-\nu_\tau)$, suggested by SuperK], (v)
the intricate pattern of the masses and mixings of the fermions,
including the smallness of $V_{cb}$ and the largeness of
$\theta^{osc}_{\nu_{\mu}\nu_{\tau}}$, and (vi) the need for B--L
as a generator to implement baryogenesis (via leptogenesis).  All
these pieces fit beautifully together within a single puzzle board
framed by supersymmetric unification, based on either SO(10) or a
string-unified G(224)-symmetry.  The two notable pieces of the
puzzle still missing, however, are proton decay and supersymmetry.

A concrete proposal is presented within a predictive
SO(10)/G(224)-framework that successfully describes the masses and
mixings of all fermions, including the neutrinos---with eight
predictions, all in agreement with observation. Within this
framework, a systematic study of proton decay is carried out,
which (a) pays special attention to its dependence on the fermion
masses, and (b) limits the threshold corrections so as to preserve
natural coupling unification. The study updates prior work by
Babu, Pati and Wilczek, in the context of both MSSM and its
(interesting) variant, the so-called ESSM, by allowing for
improved values of the matrix elements and of the short- and
long-distance renormalization effects.  It shows that a
conservative upper limit on the proton lifetime is about (1/3 -
2)$\times10^{34}$ years, with $\overline{\nu}K^{+}$ being the
dominant decay mode, and quite possibly $\mu^{+}K^{0}$ and
$e^+\pi^0$ being prominent.  This in turn strongly suggests that
an improvement in the current sensitivity by a factor of five to
ten (compared to SuperK) ought to reveal proton decay. Otherwise
some promising and remarkably successful ideas on unification
would suffer a major setback.  For comparison, some alternatives
to the conventional approach to unification pursued here are
mentioned at the end.

\end{abstract}

\vfill
\newpage

\section{\large Introduction}
\label{Introduction}

The standard model of particle physics, based on the gauge
symmetry SU(2)$_L\times $U(1)$_Y\times$ SU(3)$_C$
\cite{Weinberg,QCD} is in excellent agreement with observations,
at least up to energies of order 100 GeV.  Its success in turn
constitutes a triumph of quantum field theory, especially of the
notions of gauge invariance, spontaneous symmetry breaking, and
renormalizability.  The next step in the unification-ladder is
associated with the concept of ``grand unification'', which
proposes a unity of quarks and leptons, and simultaneously of
their three basic forces:  weak, electromagnetic and strong
\cite{JCPAS,GG,GQW}.  This concept was introduced on purely
aesthetic grounds, in fact {\it before} any of the empirical
successes of the standard model was in place.  It was realized in
1972 that the standard model judged on aesthetic merits has some
major shortcomings \cite{JCPAS,GG}.  For example, it puts members
of a family into five scattered multiplets, assigning rather
peculiar hypercharge quantum numbers to each of them, without
however providing a compelling reason for doing so.  It also does
not provide a fundamental reason for the quantization of electric
charge, and it does not explain why the electron and proton
possess exactly equal but opposite charges.  Nor does it explain
the co-existence of quarks and leptons, and that of the three
gauge forces---weak, electromagnetic and strong---with their
differing strengths.

The idea of grand unification was postulated precisely to remove these
shortcomings.  It introduces the notion that quarks and leptons are
members of one family, linked together by a symmetry group G, and that
the weak, electromagnetic and strong interactions are aspects of one
force, generated by gauging this symmetry G. The group G of course
inevitably contains the standard model symmetry $G(213)=SU(2)_L\times
U(1)_Y\times SU(3)_C$ as a subgroup.  Within this picture, the observed
differences between quarks and leptons and those between the three gauge
forces are assumed to be low-energy phenomena that arise through a
spontaneous breaking of the unification symmetry G to the standard model
symmetry G(213), at a very high energy scale $M\gg$ 1 TeV.  As a {\it
prediction} of the hypothesis, such differences must then disappear and
the true unity of quarks and leptons and of the three gauge forces
should manifest at energies exceeding the scale M.

The second and perhaps the most dramatic prediction of grand
unification is proton decay.  This important process, which would
provide the window to view physics at truly short distances ($<
10^{-30}$ cm), is yet to be seen.  Nevertheless, as I will stress
in this talk, there has appeared over the years an impressive set
of facts, favoring the hypothesis of grand unification which in
turn suggest that the discovery of proton decay should be
imminent.  These include:

{\bf (a)} {\bf The observed family structure:} The five scattered
multiplets of the standard model, belonging to a family, neatly
become parts of a whole ({\it a single multiplet}), with their
weak hypercharges predicted by grand unification, precisely as
observed.  It is hard to believe that this is just an accident.
Realization of this feature calls for an extension of the standard
model symmetry
G(213)$\,=\,$SU(2$)_{L}\times$U(1$)_{Y}\times$SU(3$)^{C}$ {\it
minimally} to the symmetry group
G(224)$\,=\,$SU(2$)_{L}\times$SU(2$)_{R}\times$ SU(4$)^{C}$
\cite{JCPAS}, which can be extended further into the simple group
SO(10) \cite{SO(10)}, but not SU(5) \cite{GG}.  The G(224)
symmetry in turn introduces some additional attractive features
(see Section \ref{Advantages}), including especially the
right-handed (RH) neutrinos ($\nu_{R}$'s) accompanying the
left-handed ones ($\nu_{L}$'s), and B--L as a local symmetry. As
we will see, both of these features now seem to be needed, on
empirical grounds, to understand neutrino masses and to implement
baryogenesis.

{\bf (b)} {\bf Quantization of electric charge and the fact that
\boldmath $Q_{\rm electron}=-Q_{\rm proton}$:} \unboldmath Grand
Unification provides compelling reasons for both of these facts.

{\bf (c)} {\bf Meeting of the gauge couplings:} Such a meeting is
found to occur at a scale $M_{X}\approx2\times10^{16}$ GeV, when
the three gauge couplings are extrapolated from their values
measured at LEP to higher energies, in the context of
supersymmetry \cite{Langacker}.  This dramatic phenomenon provides
a strong support in favor of the ideas of both grand unification
and supersymmetry \cite{SUSY}.  Both of these features in turn may
well emerge from a string theory \cite{String} or M-theory
\cite{MTh} (see discussion in Section \ref{Need}).

{\bf (d)} {\bf  \boldmath $\Delta m^2(\nu_\mu-\nu_{\tau})\sim
(1/20 eV)^2$: \unboldmath} The recent discovery of atmospheric
neutrino-oscillation at SuperKamiokande \cite{SuperK} suggests a
value  $\Delta m^2(\nu_\mu\nu_\tau) \sim (1/20\;{\rm eV})^2$. It
has been argued (see e.g.  Ref. \cite{PatiSuperK}) that precisely
such a magnitude of $\Delta m^2(\nu_\mu\nu_\tau)$ can be
understood very simply by utilizing the SU(4)-color relation
$m(\nu_\tau)_{\rm Dirac}\approx m_{\rm top}$ and the SUSY
unification scale $M_{X}$, noted above (See Section \ref{Mass}).

{\bf (e)} {\bf Some intriguing features of fermion masses and
mixings:} These include:  (i) the ``observed'' near equality of
the masses of the b-quark and the $\tau$-lepton at the
unification-scale (i.e. $m^{0}_{b}\approx m^{0}_{\tau}$) and (ii)
the observed largeness of the $\nu_{\mu}$-$\nu_{\tau}$ oscillation
angle ($\sin^{2}2\theta^{\mbox{\scriptsize
osc}}_{\nu_{\mu}\nu_{\tau}}\geq0.92$) \cite{SuperK}, together with
the smallness of the corresponding quark mixing parameter
$V_{cb}(\approx0.04)$ \cite{ParticleDataGroup}.  As shown in
recent work by Babu, Wilczek and me \cite{BabuWilczekPati}, it
turns out that these features and more can be understood
remarkably well (see discussion in Section \ref{Understanding})
within an economical and predictive SO(10)-framework based on a
minimal Higgs system.  The success of this framework is in large
part due simply to the group-structure of SO(10). For most
purposes, that of G(224) suffices.

{\bf (f)} {\bf Baryogenesis:} To implement baryogenesis
\cite{Sakharov} successfully, in the presence of electroweak
sphaleron effects \cite{KuzminRubakov}, which wipe out any baryon
excess generated at high temperatures in the (B--L)-conserving
mode, it has become apparent that one would need B--L as a
generator of the underlying symmetry in four dimensions, whose
spontaneous violation at high temperatures would yield, for
example, lepton asymmetry (leptogenesis).  The latter in turn is
converted to baryon-excess at lower temperatures by electroweak
sphalerons. This mechanism, it turns out, yields even
quantitatively the right magnitude for baryon excess
\cite{LeptoB}.  The need for B--L, which is a generator of
SU(4)-color, again points to the need for G(224) or SO(10) as an
effective symmetry near the unification-scale $M_{X}$.

The success of each of these six features (a)--(f) seems to be
non-trivial.  Together they make a strong case for both {\em the
conventional ideas on supersymmetric grand unification} and
simultaneously for the G(224)/SO(10)-route to such unification, as
being relevant to nature at short distances $\leq(10^{16}\
\mbox{GeV})^{-1}$, in four dimensions.\footnote{For comparison,
some alternative attempts, including those based on the ideas of
(a) large extra dimensions, and (b) unification occurring only in
higher dimensions, are mentioned briefly in Section 6 G.} However,
despite these successes, as long as proton decay remains
undiscovered, the hallmark of grand unification---that is {\it
quark-lepton transformability}---would remain unrevealed.

The relevant questions in this regard then are:  What is the
predicted range for the lifetime of the proton---in particular an
upper limit---within the empirically favored route to unification
mentioned above? What are the expected dominant decay modes within
this route?  Are these predictions compatible with current lower
limits on proton lifetime mentioned above, and if so, can they
still be tested at the existing or possible near-future detectors
for proton decay?

Fortunately, we are in a much better position to answer these
questions now, compared to a few years ago, because meanwhile we
have learnt more about the nature of grand unification, and also
there have been improved evaluations of the relevant matrix
elements and short and long-distance renormalization effects.  As
noted above (see also Section \ref{Advantages} and Section
\ref{Mass}), the neutrino masses and the meeting of the gauge
couplings together seem to select out the supersymmetric
G(224)/SO(10)-route to higher unification.  The main purpose of my
talk here will therefore be to address the questions raised above,
in the context of this route.  For the sake of comparison,
however, I will state the corresponding results for the case of
supersymmetric SU(5) as well.

My discussion will be based on a recent study of proton decay by Babu,
Wilczek and me \cite{BabuWilczekPati}, an update presented in the Erice
talk \cite{JCP_Erice}, and a subsequent update of the same as presented
here.  Relative to other analyses, this study has four distinctive
features:

{\bf (i)} It systematically takes into account the link that
exists between proton decay and the masses and mixings of all
fermions, including the neutrinos.

{\bf (ii)} In particular, in addition to the contributions from
the so-called ``standard'' $d=5$ operators \cite{Sakai} (see
Section \ref{Expectations}), it includes those from a {\it new}
set of $d=5$ operators, related to the Majorana masses of the RH
neutrinos \cite{BPW1}.  These latter are found to be generally as
important as the standard ones.

{\bf (iii)} As discussed in the Appendix, the work also restricts
GUT-scale threshold corrections, so as to preserve naturally
coupling unification, in accord with the observed values of the
three gauge couplings.

{\bf (iv)} Finally, the present update incorporates recently
improved values of the matrix elements, and the short and
long-distance renormalization effects, which significantly enhance
proton decay rate.

Each of these features turn out to be {\it crucial} to gaining a
reliable insight into the nature of proton decay.  Our study shows
that the inverse decay rate for the $\overline{\nu}K^{+}$-mode,
which is dominant, is less than about $1.2\times10^{31}$ years for
the case of MSSM embedded in minimal SUSY SU(5), and that it is
less than about $10^{33}$ years for the case of MSSM embedded in
SO(10).  These upper bounds are obtained by making generous
allowance for uncertainties in the matrix element and the
SUSY-spectrum.  Typically, the lifetime should of course be less
than these bounds.

Proton decay is studied also for the case of the extended
supersymmetric standard model (ESSM), that has been proposed a few
years ago \cite{BabuJi} on several grounds, based on the issues of
(a) an understanding of the inter-family mass-hierarchy, (b)
removing the mismatch between MSSM and string-unification scales,
and (c) dilaton-stabilization (see Section \ref{Expectations} and
the appendix). This case adds an extra pair of vector-like
families at the TeV-scale, transforming as
$\mathbf{16}+\overline{\mathbf{16}}$ of SO(10), to the MSSM
spectrum. While the case of ESSM is fully compatible with both
neutrino-counting at LEP and precision electroweak tests, it can
of course be tested directly at the LHC through a search for the
vectorlike fermions.  Our study shows that, with the inclusion of
only the ``standard" $d=5$ operators (defined in Section
\ref{Expectations}), ESSM, embedded in SO(10), can quite plausibly
lead to proton lifetimes in the range of $10^{33} - 10^{34}$
years, for nearly central values of the parameters pertaining to
the SUSY-spectrum and the matrix element. Allowing for a wide
variation of the parameters, owing to the contributions from both
the standard and the neutrino mass-related $d=5$ operators
(discussed in Section \ref{Expectations}), proton lifetime still
gets bounded above by about $2\times 10^{34}$ years, for the case
of ESSM, embedded in SO(10) or a string-unified G(224).

For either MSSM or ESSM, embedded in G(224) or SO(10), due to
contributions from the new operators, the $\mu^{+}K^{0}$-mode is
found to be prominent, with a branching ratio typically in the
range of 10-50\%.  By contrast, minimal SUSY SU(5), for which the
new operators are absent, would lead to branching ratios
$\leq10^{-3}$ for this mode. It is stressed that the
$e^+\pi^0$-mode induced by gauge boson-exchange, in either SUSY
SU(5) or SUSY SO(10), could have an inverse decay rate as short as
about $(1-2)\times 10^{34}$ years.

Thus our study of proton decay, correlated with fermion masses,
strongly suggests that discovery of proton decay should be
imminent.  Allowing for the possibility that the proton lifetime
may well be closer to the upper bound stated above, a
next-generation detector providing a net gain in sensitivity in
proton decay-searches by a factor of 5--10, compared to SuperK,
would certainly be needed not just to produce proton-decay events,
but also to clearly distinguish them from the background.  It
would of course also be essential to study the branching ratios of
certain sub-dominant but crucial decay modes, such as the
$\mu^{+}K^{0}$ and $e^+\pi^0$.  The importance of such improved
sensitivity, in the light of the successes of supersymmetric grand
unification, is emphasized at the end.

\section{\large Advantages of the Symmetry G(224) as a Step to Higher
     Unification}\label{Advantages}

\noindent As mentioned in the introduction, the hypothesis of
grand unification was introduced to remove some of the conceptual
shortcomings of the standard model (SM). To illustrate the
advantages of an early suggestion in this regard, consider the
five standard model multiplets belonging to the electron-family as
shown:
\begin{equation}
\left(\begin{array}{ccc}{u_{r}}\,\,\,\,{u_{y}}\,\,\,\,{u_{b}}
\\{d_{r}}\,\,\,\,{d_{y}}\,\,\,\,{d_{b}}\end{array}\right)^
{\frac{1}{3}}_{L}\,;\,\,
\left(\begin{array}{ccc}{u_{r}}\,\,\,\,{u_{y}}\,\,\,\,{u_{b}}
\end{array}\right)^{\frac{4}{3}}_{R}\,;\,\,
\left(\begin{array}{ccc}{d_{r}}\,\,\,\,{d_{y}}\,\,\,\,{d_{b}}
\end{array}\right)^{-\,\frac{2}{3}}_{R}\,;\,\,
\left(\begin{array}{c}{\nu_{e}}\\{e^{-}}\end{array}\right)
^{-\,1}_{L}\,;\,\,
\left(e^{-}\right)^{-\,2}_{R}\,.
\label{e1}
\end{equation}
Here the superscripts denote the respective weak hypercharges
$Y_{W}$ (where $Q_{em}=I_{3L}+Y_{W}/2$) and the subscripts L and R
denote the chiralities of the respective fields. If one asks: how
one can put these five multiplets into just one multiplet, the
answer turns out to be simple and unique. As mentioned in the
introduction, the minimal extension of the SM symmetry G(213)
needed, to achieve this goal, is given by the gauge symmetry
\cite{JCPAS}:
\begin{equation}
\mbox{G(224)}\,=\,\mbox{SU(2})_{L}\times \mbox{SU(2})_{R}\times
\mbox{SU(4})^{C}\,.
\label{e2}
\end{equation}
Subject to left-right discrete symmetry ($L\leftrightarrow R$), which
is natural to G(224), all members of the electron family become parts of a
single left-right self-conjugate multiplet, consisting of:
\begin{equation}
F^{e}_{L,\,R}\,=\,\left[\begin{array}{cccc}{u_{r}}\,\,\,\,
{u_{y}}\,\,\,\,{u_{b}}\,\,\,\,{\nu_{e}}\\{d_{r}}\,\,\,\,
{d_{y}}\,\,\,\,{d_{b}}\,\,\,\,{{e}^{-}}\end{array}\right]_{L,\,R}
\ . \label{e3}
\end{equation}
The multiplets $F^{e}_{L}$ and $F^{e}_{R}$ are left-right
conjugates of each other and transform respectively as (2,1,4) and
(1,2,4) of G(224); likewise for the muon and the tau families.
Note that the symmetries SU(2$)_{L}$ and SU(2$)_{R}$ are just like
the familiar isospin symmetry, except that they operate on quarks
and well as leptons, and distinguish between left and right
chiralities. The left weak-isospin SU(2$)_{L}$ treats each column
of $F^{e}_{L}$ as a doublet; likewise SU(2$)_{R}$ for $F^{e}_{R}$.
The symmetry SU(4)-color treats each row of $F^{e}_{L}$ {\it and}
$F^{e}_{R}$ as a quartet; {\it thus lepton number is treated as
the fourth color}. Note also that postulating either SU(4)-color
or SU(2$)_{R}$ forces one to introduce a right-handed neutrino
($\nu_{R}$) for each family as a singlet of the SM symmetry. {\it
This requires that there be sixteen two-component fermions in each
family, as opposed to fifteen for the SM}. The symmetry G(224)
introduces an elegant charge formula:
\begin{equation}
Q_{em}\,=\,I_{3L}\,+\,I_{3R}\,+\,\frac{B\,-\,L}{2}
\label{e4}
\end{equation}
expressed in terms of familiar quantum numbers $I_{3L}$, $I_{3R}$
and -B--L, which applies to all forms of matter (including quarks
and leptons of all six flavors, gauge and Higgs bosons). Note that
the weak hypercharge given by
$Y_{W}/2=I_{3R}\,+\,\frac{B\,-\,L}{2}$ is now completely
determined for all members of the family. The values of $Y_{W}$
thus obtained precisely match the assignments shown in Eq.
(\ref{e1}). Quite clearly, the charges $I_{3L}$, $I_{3R}$ and
B--L, being generators respectively of SU(2$)_{L}$, SU(2$)_{R}$
and SU(4$)^{c}$, are quantized; so also then is the electric
charge $Q_{em}$.

In brief, the symmetry G(224) brings some attractive features to
particle physics.  These include:\\ (i) Unification of all 16
members of a family within one left-right self-conjugate
multiplet;\\ (ii) Quantization of electric charge, with a reason
for the fact that
$Q_{\mbox{\scriptsize{electron}}}=-Q_{\mbox{\scriptsize{proton}}}$ \\
(iii) Quark-lepton unification (through SU(4) color);\\ (iv)
Conservation of parity at a fundamental level
\cite{JCPAS,RNMJCP};\\ (v) Right-handed neutrinos ($\nu_{R}'s$) as
a compelling feature; and\\ (vi) B--L as a local symmetry.\\ As
mentioned in the introduction, the two distinguishing features of
G(224)---i.e. the existence of the RH neutrinos and B--L as a
local symmetry---now seem to be needed on empirical grounds.
Furthermore, SU(4)-color provides simple relations between the
masses and mixings of quarks and leptons, while SU(2)$_L \times$
SU(2)$_R$ relates the mass-matrices in the up and down sectors. As
we will see in Sections \ref{Mass} and \ref{Understanding}, these
relations are in good accord with observations.

Believing in a complete unification, one is led to view the G(224)
symmetry as part of a bigger symmetry, which itself may have its
origin in an underlying theory, such as string theory.  In this
context, one may ask:  Could the effective symmetry below the
string scale in four dimensions (see Section \ref{Need}) be as
small as just the SM symmetry G(213), even though the latter may
have its origin in a bigger symmetry, which lives only in higher
dimensions?  I will argue in Section \ref{Mass} that the data on
neutrino masses and the need for baryogenesis provide an answer to
the contrary, suggesting that it is the {\em effective symmetry in
four dimensions, below the string scale, which must  minimally
contain either} G(224) {\em or a close relative}
G(214)$\,=\,$SU(2$)_{L}\times$I$_{3R}\times$SU(4$)^{C}$.

One may also ask:  does the effective four dimensional symmetry
have to be any bigger than G(224) near the string scale?  In
preparation for an answer to this question, let us recall that the
smallest simple group that contains the SM symmetry G(213) is
SU(5) \cite{GG}.  It has the virtue of demonstrating how the main
ideas of grand unification, including unification of the gauge
couplings, can be realized.  However, SU(5) does not contain
G(224) as a subgroup.  As such, it does not possess some of the
advantages listed above.  In particular, it does not contain the
RH neutrinos as a compelling feature, and  B--L as a local
symmetry.  Furthermore, it splits members of a family (not
including $\nu_R$) into two multiplets:
$\overline{\mathbf{5}}+\mathbf{10}$.

By contrast, the symmetry SO(10) has the merit, relative to SU(5),
that it contains G(224) as a subgroup, and thereby retains all the
advantages of G(224) listed above.  (As a historical note, it is
worth mentioning that these advantages had been motivated on
aesthetic grounds through the symmetry G(224) \cite{JCPAS}, and
{\it all} the ideas of higher unification were in place
\cite{JCPAS,GG,GQW}, before it was noted that G(224) [isomorphic
to SO(4)$\times$SO(6)] embeds nicely into SO(10) \cite{SO(10)}).
Now, {\it SO(10) even preserves the 16-plet family-structure of
G(224) without a need for any extension}.  By contrast, if one
extends G(224) to the still higher symmetry E$_{6}$ \cite{E6}, the
advantages (i)--(vi) are retained, but in this case, one must
extend the family-structure from a 16 to a 27-plet, by postulating
additional fermions.  In this sense, there seems to be some
advantage in having the effective symmetry below the string scale
to be minimally G(224) [or G(214)] and maximally no more than
SO(10).  I will compare the relative advantage of having either a
string-derived G(224) or a string-SO(10), in the next section.
First, I discuss the implications of the data on coupling
unification.

\section{\large The Need for Supersymmetry: MSSM versus String Unifications}
\label{Need}

It has been known for some time that the precision measurements of
the standard model coupling constants (in particular
$\sin^{2}\theta_{W}$) at LEP put severe constraints on the idea of
grand unification.  Owing to these constraints, the
non-supersymmetric minimal SU(5), and for similar reasons, the
one-step breaking minimal non-supersymmetric SO(10)-model as well,
are now excluded \cite{LangPolonski}.  But the situation changes
radically if one assumes that the standard model is replaced by
the minimal supersymmetric standard model (MSSM), above a
threshold of about 1 TeV.  In this case, the three gauge couplings
are found to meet \cite{Langacker}, to a very good approximation,
barring a few percent discrepancy which can be attributed to
threshold corrections (see Appendix).  Their scale of meeting is
given by \begin{equation} M_{X}\approx
2\times10^{16}\,\mbox{GeV\,\,\,\,(MSSM or SUSY\,\,SU(5))} \
.\label{e5} \end{equation}

This dramatic meeting of the three gauge couplings, or
equivalently the agreement of the MSSM-based prediction of
$\sin^{2}\theta_{W}(m_{Z})_{\rm th}=0.2315\pm0.003$ \cite{no26}
with the observed value of \break
$\sin^{2}\theta_{W}(m_{Z})=0.23124\pm0.00017$
\cite{ParticleDataGroup}, provides a strong support for the ideas
of both grand unification and supersymmetry, as being relevant to
physics at short distances $\lsim (10^{16}\, \mbox{GeV})^{-1}$.

In addition to being needed for achieving coupling unification
there is of course an independent motivation for low-energy
supersymmetry---i.e. for the existence of SUSY partners of the
standard model particles with masses of order 1 TeV.  This is
because it protects the Higgs boson mass from getting large
quantum corrections, which would (otherwise) arise from grand
unification and Planck scale physics.  It thereby provides at
least a technical resolution of the so-called gauge-hierarchy
problem. {\it In this sense low-energy supersymmetry seems to be
needed for the consistency of the hypothesis of grand
unification.} Supersymmetry is of course also needed for the
consistency of string theory.  Last but not least, as a symmetry
linking bosons and fermions, it is simply a beautiful idea. And it
is fortunate that low-energy supersymmetry can be tested at the
LHC, and possibly at the Tevatron, and the proposed NLC.

The most straightforward interpretation of the observed meeting of the
three gauge couplings and of the scale $M_{X}$, is that a supersymmetric
grand unification symmetry (often called GUT symmetry), like SU(5) or
SO(10), breaks spontaneously at $M_{X}$ into the standard model symmetry
G(213), and that supersymmetry-breaking induces soft masses of order one
TeV.

Even if supersymmetric grand unification may well be a good
effective theory below a certain scale $M \gsim M_X$, it ought to
have its origin within an underlying theory like the string/M
theory.  Such a theory is needed to unify all the forces of nature
including gravity, and to provide a good quantum theory of
gravity.  It is also needed to provide a rationale for the
existence of flavor symmetries (not available within grand
unification), which distinguish between the three families and can
resolve certain naturalness problems including those associated
with inter-family mass hierarchy.  In the context of string or
M-theory, an alternative interpretation of the observed meeting of
the gauge couplings is however possible.  This is because, even if
the effective symmetry in four dimensions emerging from a higher
dimensional string theory is non-simple, like G(224) or even
G(213), string theory can still ensure familiar unification of the
gauge couplings at the string scale. In this case, however, one
needs to account for the small mismatch between the MSSM
unification scale $M_{X}$ (given above), and the string
unification scale, given by $M_{st}\approx
g_{st}\times5.2\times10^{17}$ GeV $\approx 3.6\times10^{17}$ GeV
(Here we have put
$\alpha_{st}=\alpha_{GUT}(\mbox{MSSM})\approx0.04$)
\cite{GinsporKap}. Possible resolutions of this mismatch have been
proposed.  These include:  (i) utilizing the idea of {\it
string-duality} \cite{WittenDual} which allows a lowering of
$M_{st}$ compared to the value shown above, or alternatively (ii)
the idea of the so-called ``Extended Supersymmetric Standard
Model" (ESSM) that assumes the existence of two vector-like
families, transforming as $(\mathbf{16}+\overline{\mathbf{16}})$
of SO(10), with masses of order one TeV \cite{BabuJi}, in addition
to the three chiral families.  The latter leads to a
semi-perturbative unification by raising $\alpha_{GUT}$ to about
0.25-0.3. Simultaneously, it raises $M_{X}$, in two loop, to about
$(1/2-2)\times10^{17}$ GeV.  (Other mechanisms resolving the
mismatch are reviewed in Ref. \cite{DienesJCP}). In practice, a
combination of the two mechanisms mentioned above may well be
relevant.  \footnote{ I have in mind the possibility of
string-duality \cite{WittenDual} lowering $M_{st}$ for the case of
semi-perturbative unification in ESSM (for which
$\alpha_{st}\approx 0.25$, and thus, without the use of
string-duality, $M_{st}$ would have been about $10^{18}$ GeV) to a
value of about (1-2)$\times10^{17}$ GeV (say), and
semi-perturbative unification \cite{BabuJi} raising the MSSM value
of $M_{X}$ to about 5$\times10^{16}$ GeV$ \approx$ $M_{st}$(1/2 to
1/4) (say).  In this case, an intermediate symmetry like G(224)
emerging at $M_{st}$ would be effective only within the short gap
between $M_{st}$ and $M_{X}$, where it would break into G(213).
Despite this short gap, one would still have the benefits of
SU(4)-color that are needed to understand neutrino masses (see
Section 4), and to implement baryogenesis via leptogenesis.  At
the same time, since the gap is so small, the couplings of G(224),
unified at $M_{st}$ would remain essentially so at $M_{X}$, so as
to match with the ``observed'' coupling unification, of the type
suggested in Ref. \cite{BabuJi}.  }

While the mismatch can thus quite plausibly be removed for a
non-GUT string-derived symmetry like G(224) or G(213), a GUT
symmetry like SU(5) or SO(10) would have an advantage in this
regard because it would keep the gauge couplings together between
$M_{st}$ and $M_{X}$ (even if $M_{X}\sim M_{st}/20$), and thus not
even encounter the problem of a mismatch between the two scales. A
supersymmetric four dimensional GUT-solution [like SU(5) or
SO(10)], however, has a possible disadvantage as well, because it
needs certain color triplets to become superheavy by the so-called
doublet-triplet splitting mechanism (see Section
\ref{Expectations} and Appendix), in order to avoid the problem of
rapid proton decay.  However, no such mechanism has emerged yet,
in string theory, for the GUT-like solutions \cite{StringGUT}.
\footnote{Some alternative mechanisms for doublet-triplet
splitting, and for suppression of the $d=5$ proton decay operators
have been proposed in the context of higher dimensional theories.
These will be mentioned briefly in Section 6 G.}

Non-GUT string solutions, based on symmetries like G(224) or G(2113) for
example, have a distinct advantage in this regard, in that the dangerous
color triplets, which would induce rapid proton decay, are often
naturally projected out for such solutions
\cite{Antoniadis,FaraggiHalyo}.  Furthermore, the non-GUT solutions
invariably possess new ``flavor'' gauge symmetries, which distinguish
between families.  These symmetries are immensely helpful in explaining
qualitatively the observed fermion mass-hierarchy (see e.g.  Ref.
\cite{FaraggiHalyo}) and resolving the so-called naturalness problems of
supersymmetry such as those pertaining to the issues of
squark-degeneracy \cite{FaraggiJCP}, CP violation \cite{BabuJCP} and
quantum gravity-induced rapid proton decay \cite{JCPProton}.

Weighing the advantages and possible disadvantages of both, it
seems hard at present to make a priori a clear choice between a
GUT versus a non-GUT string-solution.  As expressed elsewhere
\cite{JCPRef}, it therefore seems prudent to keep both options
open and pursue their phenomenological consequences.  Given the
advantages of G(224) or SO(10) in the light of the neutrino masses
(see Sections \ref{Advantages} and \ref{Mass}), I will thus
proceed by assuming that either a suitable four dimensional
G(224)-solution [with the scale $M_X$ being close to $M_{st}$ (see
footnote 2)], or a realistic four-dimensional SO(10)-solution
(with the desired mechanism for doublet-triplet splitting) emerges
effectively from an underlying string theory, at the
``conventional" string-scale $M_{st}\sim 10^{17}$-$10^{18}$ GeV,
and that the G(224)/SO(10) symmetry in turn breaks spontaneously
at the conventional GUT-scale of $M_X\sim 2\times 10^{16}$ GeV (or
at $M_X\sim 5\times 10^{16}$ GeV for the case of ESSM, as
discussed in footnote 2) to the standard model symmetry G(213).
The extra dimensions of string/M-theory are assumed to be tiny
with sizes $\leq M_X^{-1}\sim 10^{-30}$ cm, so as not to disturb
the successes of GUT.  In short, I assume that essentially {\it
the conventional (good old) picture of grand unification, proposed
and developed sometime ago \cite{JCPAS,GG,GQW,SO(10),Langacker},
holds as a good effective theory above the unification scale $M_X$
and up to some high scale $M\lsim M_{st}$, with the added
presumption that it may have its origin from the string/M-theory.}
Such a picture seems to be directly motivated on observational
grounds such as those based on (a) coupling unification (discussed
above), (b) neutrino masses including the (mass)$^2$-difference of
the $\nu_\mu$-$\nu_\tau$ system and the near maximal
$\nu_\mu$-$\nu_\tau$ oscillation angle (see discussions in the
next sections), and (c) the fact that spontaneous violation of
B--L local symmetry at high temperatures, seems to be needed to
implement baryogenesis via leptogenesis.\footnote{Alternative
scenarios such as those based on TeV-scale large extra dimensions
\cite{ref:36} or string-scale being at a few TeV \cite{ref:37}, or
submillimeter-size even larger extra dimensions with the
fundamental scale of quantum gravity being a few TeV
\cite{largeextradim}, though intriguing, do not seem to provide
simple explanations of these features: (a), (b) and (c). They will
be mentioned briefly in Section  6 G.}

 We will see that with the broad assumption mentioned above, an economical
 and predictive framework emerges, which successfully accounts for a host of
 observed phenomena pertaining to the masses and the mixings of all fermions,
including neutrinos. It also makes some crucial testable predictions for
proton decay. I next discuss the implications of the mass of $\nu_\tau$, or
rather of $\Delta m^2(\nu_\mu\nu_\tau)$, as revealed by the SuperK data.

\section{ \large \boldmath $\Delta m^2(\nu_\mu\nu_\tau)$: \unboldmath
Evidence In Favor of the G(224) Route}
\label{Mass}

One can obtain an estimate for the mass of $\nu_{L}^{\tau}$ in the
context of G(224) or SO(10) by using the following three steps
(see e.g. Ref. \cite{PatiSuperK}):

(i) Assume that B$-$L and $I_{3R}$, contained in a string-derived
G(224) or SO(10), break near the unification-scale:
\begin{equation}
M_X\sim2\times10^{16}\,\mbox {GeV}\,,
\label{e6}
\end{equation}
through VEVs of Higgs multiplets of the type suggested by
string-solutions---i.e. $\langle(1, 2, 4)_H\rangle$ for G(224) or
$\langle\overline{\mathbf{16}}_{H}\rangle$ for  SO(10), as opposed
to $\mathbf{126}_H$ which seems to be unobtainable at least in
weakly interacting string theory \cite{DienesRussell}. In the
process, the RH neutrinos  ($\nu_{R}^{i}$), which are singlets of
the standard model, can and  generically will acquire superheavy
Majorana masses of the type
$M_{R}^{ij}\,\nu_R^{iT}\,C^{-1}\,\nu_R^{j}$, by utilizing the VEV
of $\langle\overline{\mathbf{16}}_{H}\rangle$ and effective
couplings of the form:
\begin{equation}
{\cal
L}_M\,(SO(10))\,=\,f_{ij}\,\,\mathbf{16}_{i}\cdot\mathbf{16}_{j}\,\,
\overline\mathbf{16}_{H}\,\cdot\overline{\mathbf{16}}_{H}/M + h.c.
\label{e7}
\end{equation}

A similar expression holds for G(224).  Here $i,j=1,2,3$,
correspond respectively to $e,\,\mu$ and $\tau$ families.  Such
gauge-invariant non-renormalizable couplings might be expected to
be induced by Planck-scale physics, involving quantum gravity or
stringy effects and/or tree-level exchange of superheavy states,
such as those in the string tower.  With $f_{ij}$ (at least the
largest among them) being of order unity, we would thus expect M
to lie between $M_{\rm Planck}\approx2\times10^{18}$ GeV and
$M_{\rm string}\approx4\times10^{17}$ GeV.  Ignoring for the
present off-diagonal mixings (for simplicity), one thus obtains
\footnote{The effects of neutrino-mixing and of the more
legitimate choice of $M=M_{string}\approx4\times 10^{17}$ GeV
(instead of $M=M_{\rm Planck}$) on the values of $m(\nu^\tau_L)$
and of $M_{3R}$ are considered in Ref. \cite{BabuWilczekPati} and
are reflected in our discussions in Section 5. The two effects
together end up in yielding essentially the same mass for
$m(\nu_L^\tau)$ as obtained within the simplified picture
presented in this section, together with a value for
$M_{3R}\approx(5$-$10)\times 10^{14}$ GeV.}:
\begin{equation}
M_{3R}\,\approx\,\frac{f_{33}\langle\overline{\mathbf{16}}_{H}\rangle^{2}}{M}
\,\approx\,f_{33}\,(2\times10^{14}\,\mbox{GeV})\,\rho^{2}\,(M_{\rm
Planck}/M) \label{e8}
\end{equation}

This is the Majorana mass of the RH tau neutrino.  Guided by the
value of $M_{X}$, we have substituted
$\langle\overline{\mathbf{16}}_{H}\rangle=(2\times10^{16}\,\mbox{GeV})\,\rho$
,where we expect $\rho\approx1/2$ to 2 (say).

(ii) Now using SU(4)-color and the Higgs multiplet
$(\mathbf{2},\mathbf{2},\mathbf{1})_H$ of G(224) or equivalently
$\mathbf{10}_{H}$ of SO(10), one obtains the relation
$m_{\tau}(M_{X}) = m_{b}(M_{X})$, which is known to be successful.
Thus, there is a good reason to believe that the third family gets
its masses primarily from the $\mathbf{10}_{H}$ or equivalently
$(2, 2, 1)_H$ (see Section 5).  In turn, this implies:
\begin{equation} m(\nu^{\tau}_{\rm
Dirac})\,\approx\,m_{\rm top}(M_{X})\,\approx\,(100\,
\mbox{-}\,120)\,\mbox{GeV}\ . \label{e9}
\end{equation} Note that
this relationship between the Dirac mass of the tau-neutrino and
the top-mass is special to SU(4)-color.  It does not emerge in
SU(5).

(iii) Given the superheavy Majorana masses of the RH neutrinos as
well as the Dirac masses as above, the see-saw mechanism
\cite{SeeSaw} yields naturally light masses for the LH neutrinos.
For $\nu_{L}^{\tau}$ (ignoring flavor-mixing), one thus obtains,
using Eqs.(\ref{e8}) and (\ref{e9}),
\begin{equation}
m(\nu^{\tau}_{L})\,\approx\,\frac{m(\nu^{\tau}_{\rm Dirac})^2}
{M_{3R}}\,\approx\,[(1/20)\,\mbox{eV}\,(1\,\mbox{-}\,1.44)
/f_{33}\,\rho^{2}]\,(M/M_{\rm Planck})\ . \label{e10}
\end{equation}

In the next section, we discuss the masses and mixings of all
three neutrinos.  As we will see, given the hierarchical masses of
quarks and charged leptons and the see-saw mechanism, we naturally
obtain $m(\nu_L^\mu)\sim (1/10)m(\nu_L^\tau)$.  We are thus led to
predict that $\Delta m^2(\nu_\mu\nu_\tau)_{th}\equiv
|m^2(\nu^\tau_L)-m^2(\nu^\mu_L)|_{th} \approx
m^2(\nu^\tau_L)_{th}=\mbox{ square of the RHS of Eq.
(\ref{e10})}$. Now SuperK result strongly suggests that it is
observing $\nu_L^\mu$-$\nu_L^\tau$ (rather than
$\nu_L^\mu$-$\nu_X$) oscillation, with a \break
 $\Delta
m^2(\nu_\mu\nu_\tau)_{obs}\approx 3\times 10^{-3}$ eV$^2$.  It
seems {\it truly remarkable} that the expected magnitude of
$\Delta m^2(\nu_\mu\nu_\tau)$, given to a very good approximation
by the square of the RHS of Eq.  (\ref{e10}), is just about what
is observed at SuperK, if $f_{33}\,\rho^2(M_{\rm
Planck}/M)\approx$ 1.3 to 1/2.  Such a range for
$f_{33}\,\rho^2(M_{\rm Planck}/M)$ seems most plausible and
natural (see discussion in Ref. \cite{PatiSuperK}).  Note that the
estimate (\ref{e10}) crucially depends upon the supersymmetric
unification scale, which provides a value for $M_{3R}$, as well as
on SU(4)-color that yields $m(\nu^{\tau}_{\rm Dirac})$. {\it The
agreement between the expected and the SuperK results thus clearly
favors supersymmetric unification, and in the string theory
context, it suggests that the effective symmetry below the
string-scale should contain SU(4)-color}.  Thus, minimally this
effective symmetry should be either G(214) or G(224), and
maximally as big as SO(10), if not E$_{6}$.

By contrast, if SU(5) is regarded as either a fundamental symmetry
or as the effective symmetry below the string scale, there would
be no compelling reason based on symmetry alone, to introduce a
$\nu_{R}$, because it is a singlet of SU(5).  Second, even if one
did introduce $\nu^{i}_{R}$ by hand, their Dirac masses, arising
from the coupling $h^{i}\,\overline\mathbf{5}_{i}\langle
\mathbf{5}_H\rangle\nu^{i}_{R}$, would be unrelated to the
up-flavor masses and thus rather arbitrary [contrast with Eq.
(\ref{e9})]. So also would be the Majorana masses of the
$\nu^{i}_{R}$'s, which are SU(5)-invariant, and thus can be even
of order string scale . This would give extremely small values of
$m(\nu_L^\tau)$ and $m(\nu_L^\mu)$ and thus of $\Delta
m^2(\nu_\mu\nu_\tau)$, which would be in gross conflict with
observation.

Before passing to the next section, it is worth noting that the
mass of $\nu_{\tau}$ or of $\Delta m^2(\nu_\mu \nu_\tau)$
suggested by SuperK, as well as the observed value of
$\sin^{2}\theta_{W}$ (see Section \ref{Need}), provide valuable
insight into the nature of GUT symmetry breaking. They both favor
the case of a {\it single-step breaking} (SSB) of SO(10) or a
string-unified G(224) symmetry at a high scale of order $M_{X}$,
into the standard model symmetry G(213), as opposed to that of a
multi-step breaking (MSB).  The latter would correspond, for
example, to SO(10) [or G(224)] breaking at a scale $M_{1}$ into
G(213), which in turn breaks at a scale $M_{2}\ll M_{1}$ into
G(213).  One reason why the case of single-step breaking is
favored over that of MSB is that the latter can accommodate but
not really predict $\sin^{2}\theta_{W}$, whereas the former
predicts the same successfully.  Furthermore, since the Majorana
mass of $\nu^{\tau}_{R}$ arises arises only after B--L and
$I_{3R}$ break, it would be given, for the case of MSB, by
$M_{3R}\sim f_{33}(M_{2}^{2}/M)$, where $M\sim M_{st}$ (say).  If
$M_{2}\ll M_{X}\sim 2\times10^{16}$ GeV, and M $>M_{X}$, one would
obtain too low a value ($\ll 10^{14}$ GeV) for $M_{3R}$ [compare
with Eq. (8)], and thereby too large a value for
$m(\nu^{\tau}_{L})$, compared to that suggested by SuperK.  By
contrast, the case of single-step breaking (SSB) yields the right
magnitude for $m(\nu_{\tau})$ [see Eq.  (10)].

{\it Thus the success of the results on $m(\nu_{\tau})$ and
thereby on $\Delta m^2(\nu_\mu\nu_\tau)$ discussed above not only
favors the symmetry SO(10) or G(224) beging effective in 4D at a
high scale, but also clearly suggests that B--L and $I_{3R}$ break
near the conventional GUT scale $M_{X}\sim 2\times 10^{16}$ GeV,
rather than at an intermediate scale $\ll M_{X}$.} In other words,
the observed values of both $\sin^{2}\theta_{W}$ and $\Delta
m^2(\nu_\mu\nu_\tau)$ favor only {\it the simplest pattern of
symmetry-breaking}, for which SO(10) or a string-derived G(224)
symmetry breaks in one step to the standard model symmetry, rather
than in multiple steps.  It is of course only this simple pattern
of symmetry breaking that would be rather restrictive as regards
its predictions for proton decay (to be discussed in Section
\ref{Expectations}).  I next discuss the problem of understanding
the masses and mixings of all fermions.

\section{\large Understanding Aspects of Fermion Masses and
Neutrino Oscillations in SO(10)}
\label{Understanding}

Understanding the masses and mixings of all quarks {\it in
conjunction with} those of the charged leptons {\it and} neutrinos
is a goal worth achieving by itself.  It also turns out to be
essential for the study of proton decay.  I therefore present
first a partial attempt in this direction, based on a quark-lepton
unified G(224)/SO(10)-framework, which seems most promising
\cite{BabuWilczekPati}.  A few guidelines would prove to be
helpful in this regard.  The first of these is motivated by the
desire for economy [see (\ref{e11})], and the rest (see below) by
the data.  In essence, we will be following (partly) a {\em
bottom-up approach} by appealing to the data to provide certain
clues as regards the pattern of the Yukawa couplings, and
simultaneously a {\em top-down approach} by appealing to grand
unification,  based on the symmetry G(224)/SO(10), to restrict the
couplings by the constraints of group theory.  The latter helps to
interrelate the masses and mixings of quarks with those of the
charged leptons and the neutrinos.  As we will see, it is these
{\it interrelationships}, which permit predictivity, and are found
to be remarkably successful.  The guidelines which we adopt are as
follows.

{\bf 1) Hierarchy Through Off-diagonal Mixings:} Recall earlier
attempts \cite{Weinbergetal} that attribute hierarchical masses of
the first two families to mass matrices of the form:
\begin{equation}
M\,=\,\left(\begin{array}{cc}{0}\,\,\,\,\,\,\,\,\,\,\,\,{\epsilon}
\\{\epsilon}\,\,\,\,\,\,\,\,\,\,\,\,{1}\end{array}\right)\,m^{(0)}_{s}\,,
\label{e11} \end{equation} for the $(d,s)$ quarks, and likewise
for the $(u,c)$ quarks.  Here $\epsilon\sim1/10$.  The
hierarchical patterns in Eq.  (\ref{e11}) can be ensured by
imposing a suitable flavor symmetry which distinguishes between
the two families (that in turn may have its origin in string
theory (see e.g.  Ref \cite{FaraggiHalyo}).  Such a pattern has
the virtues that (a) it yields a hierarchy that is much larger
than the input parameter $\epsilon$:
$(m_{d}/m_{s})\approx\epsilon^{2}\ll\epsilon$, and (b) it leads to
an expression for the Cabibbo angle:  \begin{equation}
\theta_{c}\approx\bigg|\sqrt{\frac{m_{d}}{m_{s}}}\,-\,e^{i\phi}\,
\sqrt{\frac{m_{u}}{m_{c}}}\,\bigg|\,, \label{e12} \end{equation}
which is rather successful.  Using $\sqrt{m_{d}/m_{s}}\approx
0.22$ and $\sqrt{m_{u}/m_{c}}\approx 0.06$, we see that Eq.
(\ref{e12}) works to within about $25\%$ for any value of the
phase $\phi$.  Note that the square root formula (like
$\sqrt{m_{d}/m_{s}}$) for the relevant mixing angle arises because
of the symmetric form of $M$ in Eq.  (\ref{e11}), which in turn is
ensured if the contributing Higgs is a 10 of SO(10).  A
generalization of the pattern in Eq.  (\ref{e11}) would suggest
that the first two families (i.e. the $e$ and the $\mu$) receive
masses primarily through their mixing with the third family
$(\tau)$, with $(1,3)$ and $(1,2)$ elements being smaller than the
$(2,3)$; while $(2,3)$ is smaller than the $(3,3)$.  We will
follow this guideline, except for the modification noted below.

{\bf 2) The Need for an Antisymmetric Component:} Although the
symmetric hierarchical matrix in Eq.  (\ref{e11}) works well for
the first two families, a matrix of the same form fails altogether
to reproduce $V_{cb}$, for which it yields:
\begin{equation}
V_{cb}\approx\bigg|\sqrt{\frac{m_{s}}{m_{b}}}\,-\,e^{i\chi}\,
\sqrt{\frac{m_{c}}{m_{t}}}\,\bigg|\,. \label{e13}
\end{equation}
Given that $\sqrt{m_{s}/m_{b}}\approx 0.17$ and
$\sqrt{m_{c}/m_{t}}\approx 0.0.06$, we see that Eq.  (\ref{e13})
would yield $V_{cb}$ varying between 0.11 and 0.23, depending upon
the phase $\chi$.  This is too big, compared to the observed value
of $V_{cb}\approx0.04\pm0.003$, by at least a factor of 3. We
interpret this failure as a {\it clue} to the presence of an
antisymmetric component in $M$, together with symmetrical ones (so
that $m_{ij}\neq m_{ji}$), which would modify the relevant mixing
angle to $\sqrt{m_i/m_j}\ \sqrt{m_{ij}/m_{ji}}$, where $m_{i}$ and
$m_{j}$ denote the respective eigenvalues.

{\bf 3) The Need for a Contribution Proportional to B--L:} The
success of the relations $m^{0}_{b}\approx m^{0}_{\tau}$, and
$m^{0}_{t}\approx m(\nu_{\tau})^{0}_{\rm Dirac}$ (see Section
\ref{Mass}), suggests that the members of the third family get
their masses primarily from the VEV of a SU(4)-color singlet Higgs
field that is independent of B--L.  This is in fact ensured if the
Higgs is a 10 of SO(10). However, the empirical observations of
$m^{0}_{s}\sim m^{0}_{\mu}/3$ and $m^{0}_{d}\sim 3m^{0}_{e}$
\cite{GeorgiJarsklog} call for a contribution proportional to B--L
as well.  Further, one can in fact argue that understanding
naturally the suppression of $V_{cb}$ (in the quark-sector)
together with an enhancement of $\theta^{\rm
osc}_{\nu_{\mu}\,\nu_{\tau}}$ (in the lepton sector) calls for a
contribution that is not only proportional to B--L, but also
antisymmetric in the family space (this later feature is suggested
already in item (\ref{e2})).  We show below how both of these
requirements can be met in SO(10), even for a minimal Higgs
system.

{\bf 4) Up-Down Asymmetry:} Finally, the up and the down-sector
mass matrices must not be proportional to each other, as otherwise
the CKM angles would all vanish.  Note that the cubic couplings of
a single $10_H$ with the fermions in the 16's will not serve the
purpose in this regard.

Following Ref. \cite{BabuWilczekPati}, I now present a simple and
predictive mass-matrix, based on SO(10), that satisfies {\it all
four} requirements (\ref{e1}), (\ref{e2}), (\ref{e3}) and
(\ref{e4}). The interesting point is that one can obtain such a
mass-matrix for the fermions by utilizing only the minimal Higgs
system, that is needed anyway to break the gauge symmetry SO(10).
It consists of the set:
\begin{equation}
H_{\rm
minimal}\,=\,\{\mathbf{45}_{H},\,\mathbf{16}_{H},\,\overline{\mathbf{16}}_{H},\,
\mathbf{10}_{H}\}\,. \label{e14}
\end{equation}
Of these, the VEV of $\langle\mathbf{45}_{H}\rangle\sim M_{X}$
breaks SO(10) into G(2213), and those of
$\langle\mathbf{16}_{H}\rangle=\langle\overline\mathbf{16}_{H}\rangle\sim
M_{X}$
 break
G(2213) to G(213), at the unification-scale $M_{X}$. Now G(213) breaks
at the electroweak scale by the VEV of $\langle10_{H}\rangle$ to
U$(1)_{em}\times$ SU$(3)^{c}$.

One might have introduced large-dimensional tensorial multiplets
of SO(10) like $\overline\mathbf{126}_{H}$ and $\mathbf{120}_{H}$,
both of which possess cubic level Yukawa couplings with the
fermions. In particular, the coupling
$\mathbf{16}_{i}\mathbf{16}_{j}(\mathbf{120}_{H})$ would give the
desired family-antisymmetric as well as (B--L)-dependent
contribution. We do not however introduce these multiplets in part
because there is a general argument suggesting that they do not
arise at least in weakly interacting heterotic string solutions
\cite{DienesRussell}, and in part also because mass-splittings
within such large-dimensional multiplets could give excessive
threshold corrections to $\alpha_{3}(m_{z})$ (typically exceeding
20\%), rendering observed coupling unification fortuitous.  By
contrast, the multiplets in the minimal set (shown above) can
arise in string solutions. Furthermore, the threshold corrections
for the minimal set are found to be naturally small, and even to
have the right sign, to go with the observed coupling unification
\cite{BabuWilczekPati} (see Appendix).

The question is:  can the minimal set of Higgs multiplets [see Eq.
(\ref{e14})] meet all the requirements listed above?  Now
$\mathbf{10}_{H}$ (even several $\mathbf{10}$'s) cannot meet the
requirements of antisymmetry and $(B$-$L)$-dependence.
Furthermore, a single $\mathbf{10}_{H}$ cannot generate
CKM-mixings. This impasse disappears, however, as soon as one
allows for not only cubic, but also effective non-renormalizable
quartic couplings of the minimal set of Higgs fields with the
fermions. These latter couplings could of course well arise
through exchanges of superheavy states (e.g. those in the string
tower) involving renormalizable couplings, and/or through quantum
gravity.

Allowing for such cubic and quartic couplings and adopting the
guideline (\ref{e1}) of hierarchical Yukawa couplings, as well as
that of economy, we are led to suggest the following effective
lagrangian for generating Dirac masses and mixings of the three
families \cite{BabuWilczekPati} (for a related but different
pattern, involving a non-minimal Higgs system, see Ref.
\cite{AlbrightBarr}).
\begin{eqnarray}
\mathbf{{\cal
L}_{Yuk}}\,=\,h_{33}\,\mathbf{16}_{3}\,\mathbf{16}_{3}\,
\mathbf{10}_{H}\,+\,[\,h_{23}\,\mathbf{16}_{2}\,
\mathbf{16}_{3}\,\mathbf{10}_{H}\,+\,a_{23}\,
\mathbf{16}_{2}\,\mathbf{16}_{3}\,\mathbf{10}_{H}\,
\mathbf{45}_{H}/M\,\nonumber\\[.5ex]&&
{\hspace{-12cm}}+\,g_{23}\,\mathbf{16}_{2}\,\mathbf{16}_{3}\,
\mathbf{16}_{H}\,\mathbf{16}_{H}/M]\,+\,\{a_{12}\,
\mathbf{16}_{1}\,\mathbf{16}_{2}\,\mathbf{10}_{H}\,\mathbf{45}_{H}/M\,
\nonumber\\[.5ex]&&
{\hspace{-10.5cm}}+\,g_{12}\,\mathbf{16}_{1}\,\mathbf{16}_{2}\,
\mathbf{16}_{H}\,\mathbf{16}_{H}/M\}\,.
\label{e15}
\end{eqnarray}
Here, $M$ could plausibly be of order string scale. Note that a
mass matrix having essentially the form of Eq. (\ref{e11})
results if the first term $h_{33}\langle\mathbf{10}_{H}\rangle$ is
dominant. This ensures $m^{0}_{b}\approx m^{0}_{\tau}$ and
$m^{0}_{t}\approx m^{0}(\nu_{\rm Dirac})$. Following the
assumption of progressive hierarchy (equivalently appropriate
flavor symmetries \footnote{Although no
  explicit string solution with the hierarchy in all the Yukawa couplings
  in Eq. (\ref{e15})---i.e. in $h_{ij}$, $a_{ij}$ and $g_{ij}$---exists as yet,
  one can postulate flavor symmetries of
  the type alluded to (e.g. two abelian U(1) symmetries), which assign
  flavor charges not only to the fermion families and the Higgs multiplets,
  but also to a few (postulated) SM singlets that acquire VEVs of order
  M$_X$. The flavor symmetry-allowed effective couplings such as
  $\mathbf{16}_2 \mathbf{16}_3 \mathbf{10}_H \VEV S/M$ would lead to
  $h_{23}\sim \VEV S/M \sim 1/10$.
  One can verify that the full set of hierarchical couplings shown in
  Eq. (\ref{e15}) can in fact arise in the presence of two such U(1)
  symmetries. String theory
  (at least) offers the scope (as indicated by the solutions of
  Refs. \cite{FaraggiHalyo} and \cite{Antoniadis}) for providing a rationale
  for the existence of such flavor symmetries, together with that of the
  SM singlets. For example, there exist solutions with the
  top Yukawa coupling being leading and others being hierarchical
  (as in Ref. \cite{FaraggiHalyo}).}),
we presume that $h_{23}\sim h_{33}/10$, while $h_{22}$ and
$h_{11}$, which are not shown, are assumed to be progressively
much smaller than $h_{23}$. Since
$\langle\mathbf{45}_{H}\rangle\sim\langle\mathbf{16}_{H}\rangle\sim
M_{X}$, while $M\sim M_{st}\sim10M_{X}$, the terms
$a_{23}\langle\mathbf{45}_{H}\rangle/M$ and
$g_{23}\langle\mathbf{16}_{H}\rangle/M$ can quite plausibly be of
order $h_{33}/10$, if $a_{23}\sim g_{23}\sim h_{33}$. By the
assumption of hierarchy, we presume that $a_{12}\ll a_{23}$, and
$g_{12}\ll g_{23}$

It is interesting to observe the symmetry properties of the
$a_{23}$ and $g_{23}$-terms.  Although
$\mathbf{10}_{H}\times\mathbf{45}_{H}=\mathbf{10+120+320}$, given
that $\langle\mathbf{45}_{H}\rangle$ is along B--L, which is used
to implement doublet-triplet splitting (see Appendix), only
$\mathbf{120}$ in the decomposition contributes to the
mass-matrices. This contribution is, however, antisymmetric in the
family-index and, at the same time, proportional to B--L. {\it
Thus the $a_{23}$ term fulfills the requirements of both
antisymmetry and (B--L)-dependence, simultaneously \footnote{The
analog of $\mathbf{10}_{H}\cdot\mathbf{45}_{H}$ for the case of
G(224) would be $\chi_{H}\equiv(2,2,1)_{H}\cdot(1,1,15)_{H}$.
Although in general, the coupling of $\chi_{H}$ to the fermions
need not be antisymmetric, for a string-derived G(224), the
multiplet (1,1,15$)_{H}$ is most likely to arise from an
underlying 45 of SO(10) (rather than 210); in this case, the
couplings of $\chi_{H}$ must be antisymmetric like that of
$\mathbf{10}_{H}\cdot\mathbf{45}_{H}$.} }. With only $h_{ij}$ and
$a_{ij}$-terms, however, the up and down quark mass-matrices will
be proportional to each other, which would yield $V_{CKM} =1$.
This is remedied by the $g_{ij}$ coupling, because, the
$\mathbf{16}_{H}$ can have a VEV not only along its SM singlet
component (transforming as $\tilde{\overline{\nu}}_{R}$) which is
of GUT-scale, but also along its electroweak doublet
component---call it $\mathbf{16}_{d}$---of the electroweak scale.
The latter can arise by the the mixing of $\mathbf{16}_{d}$ with
the corresponding doublet (call it $\mathbf{10}_{d}$) in the
$\mathbf{10}_{H}$.  The MSSM doublet $H_{d}$, which is light, is
then a mixture of $\mathbf{10}_{d}$ and $\mathbf{16}_{d}$, while
the orthogonal combination is superheavy (see Appendix).  Since
$\langle\mathbf{16}_{d}\rangle$ contributes only to the
down-flavor mass matrices, but not to the up-flavor, the $g_{23}$
and $g_{12}$ couplings generate non-trivial CKM-mixings.  {\it We
thus see that the minimal Higgs system (as shown in Eq.
(\ref{e14})) satisfies {\em a priori} all the qualitative
requirements (1)--(4), including the condition of $V_{CKM}\neq1$}.
I now discuss that this system works well even quantitatively.

With the six effective Yukawa couplings shown in Eq. (\ref{e15}),
the Dirac mass matrices of quarks and leptons of the three
families at the unification scale take the form:
\begin{eqnarray}
U\ &=& \left(\begin{array}{ccc}{0} & {\epsilon'} & {0} \\
{-\,\epsilon'} & {0}  & {\epsilon\,+\,\sigma} \\ {0} &
{-\,\epsilon\,+\,\sigma} & {1}\end{array}\right)\,m_U,\nonumber
\\[1.5em]
D &=& \left(\begin{array}{ccc}{0} & {\epsilon'\,+\,\eta'} & {0} \\
{-\,\epsilon'\,+\,\eta'} & {0} & {\epsilon\,+\,\eta} \\ {0} &
{-\,\epsilon\,+\,\eta} & {1}\end{array}\right)\,m_D,  \nonumber \\[1.5em]
N &=& \left(\begin{array}{ccc}{0} & {-\,3\epsilon'} & {0} \\
{3\epsilon'} & {0} & {-\,3\epsilon\,+\,\sigma} \\ {0} &
{3\epsilon\,+\,\sigma} & {1}\end{array}\right)\,m_U,\nonumber
\\[1.5em]
L &=& \left(\begin{array}{ccc}{0} & {-\,3\epsilon'\,+\,\eta'} & {0} \\
{3\epsilon'\,+\,\eta'} & {0} & {-\,3\epsilon\,+\,\eta} \\ {0} &
{3\epsilon\,+\,\eta} & {1}\end{array}\right)\,m_D. \label{e16}
\end{eqnarray}
Here the matrices are multiplied by  left-handed fermion fields
from the left and by anti--fermion fields from  the right. $(U,D)$
stand for the mass matrices of up and  down quarks, while $(N,L)$
are the Dirac mass matrices  of the neutrinos and the charged
leptons. The entries $1,\epsilon$,and $\sigma$ arise respectively
from the $h_{33},a_{23}$ and $h_{23}$ terms in Eq. (\ref{e15}),
while $\eta$ entering into $D$ and $L$ receives contributions from
both  $g_{23}$ and $h_{23}$; thus $\eta\neq\sigma$. Similarly
$\eta'$ and $\epsilon'$ arise from $g_{12}$ and $a_{12}$ terms
respectively. Note the quark-lepton correlations between $U$ and
$N$ as well as $D$ and $L$ arise because of SU(4)$^C$, while the
up-down correlations between $U$ and $D$ as well as $N$ and $L$
arise because of SU(2)$_L\times$SU(2)$_R$. Thus, these
correlations emerge just because of the symmetry property of
G(224). The relative factor of $-3$ between  quarks and leptons
involving the $\epsilon$ entry reflects the fact  that
$\langle\mathbf{45_{H}}\rangle$ is proportional to (B--L), while
the antisymmetry in this entry arises from the group structure of
SO(10),  as explained above$^7$. As we will see, this
$\epsilon$-entry helps to account for (a) the differences between
$m_{s}$ and $m_{\mu}$, (b) that between $m_{d}$ and $m_{e}$, and
most important, (c) the suppression of $V_{cb}$ {\it together
with} the enhancement of the $\nu_{\mu}$-$\nu_{\tau}$ oscillation
angle.

The mass matrices in Eq. (\ref{e16}) contain 7 parameters
\footnote{
   Of these,  $m_{U}^{0}\approx m_{t}^{0}$ can in fact be estimated to
   within $20\%$  accuracy by either using the argument of radiative
   electroweak symmetry  breaking, or some promising string solutions
   (see e.g. Ref. \cite{FaraggiHalyo}).
}: $\epsilon$, $\sigma$, $\eta$,
$m_{D}=h_{33}\,\langle10_{d}\rangle$,
$m_{U}=h_{33}\,\langle10_{U}\rangle$, $\eta'$ and $\epsilon'$.
These may be determined by using, for  example, the following
input values: $m_{t}^{\rm phys}=174$ GeV, $m_{c}(m_{c})=1.37$ GeV,
$m_{s}(1$ GeV$)=110$--$116$ MeV \cite{Gupta}, $m_{u}(1$ GeV)
$\approx6$ MeV and the observed masses of $e$, $\mu$ and $\tau$,
which lead to (see Ref. \cite{BabuWilczekPati}, for details):
\[
\sigma\,\simeq\,\,0.110\,,\,\,\eta\simeq\,0.151\,,\,\,
\epsilon\,\simeq\,-\,0.095\,,\,\,|\eta'|\approx4.4\times10^{-3}
\,\,\,\mbox{and}\,\,\,\epsilon'\approx2\times10^{-4}
\]
\begin{equation}
m_{U}\,\simeq\,m_{t}(M_{U})\,\simeq\,(100\mbox{-}120)\,
\mbox{GeV}\,,\,\,m_{D}\,\simeq\,m_{b}(M_{U})\,\simeq\,1.5\,\mbox{GeV}\,.
\label{e17}
\end{equation}

Here, I will assume, only for the sake of simplicity, as in Ref.
\cite{BabuWilczekPati}, that the parameters are real.\footnote{
Babu and I have recently studied supersymmetric CP violation
within the G(224)/SO(10) framework, by using precisely the fermion
mass-matrices as in Eq. (\ref{e16}).  We have observed
\cite{BabuJCP} that complexification of the parameters can lead to
observed CP violation, without upsetting in the least the success
of Ref. \cite{BabuWilczekPati} (i.e. of the fermion mass-matrices
of Eq. (\ref{e16})) in describing the masses and mixings of all
fermions, including neutrinos.  Even with complexification the
relative signs and the approximate magnitudes of the real parts of
the parameters must be the same as in Eq. (\ref{e17}), to retain
the success.  } Note that in accord with our general expectations
discussed above, each of the parameters $\sigma$, $\eta$ and
$\epsilon$ are found to be of order 1/10, as opposed to being
\footnote{This is one characteristic difference between our work
and that of Ref. \cite{AlbrightBarr}, where the (2,3)-element is
even bigger than the (3,3).  } $O(1)$ or $O(10^{-2})$, compared to
the leading (3,3)-element in Eq.  (\ref{e16}).  Having determined
these parameters, we are led to a total of five predictions
involving only the quarks (those for the leptons are listed
separately):  \begin{equation}
m^{0}_{b}\,\approx\,m^{0}_{\tau}(1\,-\,8\epsilon^{2})\,;\,\,\,\,
\mbox{thus}\,\,\,\,m_{b}(m_{b})\,\simeq\,(4.6\mbox{-}4.9)\,\mbox{GeV}
\label{e18}
\end{equation}

\begin{equation}
|V_{cb}| \simeq |\sigma\,-\,\eta|\label{e19}  \approx
\left|\sqrt{m_{s}/m_{b}}\left|\frac{\eta\,+
\,\epsilon}{\eta\,-\,\epsilon}\right|^{1/2}\,-
\,\sqrt{m_{c}/m_{t}}\,\left|\frac{\sigma\,+\,\epsilon}{\sigma\,-
\,\epsilon}\right|^{1/2}\right|\,\simeq\,0.045
\end{equation}
\begin{equation} m_{d}\,(1
\mbox{GeV})\,\simeq\,8\,\mbox{MeV} \label{e20}
\end{equation}

\begin{equation}
\theta_{C}\,\simeq\,\left|\sqrt{m_{d}/m_{s}}\,-
\,e^{i\phi}\sqrt{m_{u}/m_{c}}\right| \label{e21}
\end{equation}

\begin{equation}
|V_{ub}/V_{cb}|\,\simeq\,\sqrt{m_{u}/m_{c}}\,\simeq\,0.07\,.
\label{e22}
\end{equation}
In making these predictions, we have
extrapolated the GUT-scale values down to low energies using
$\alpha_{3}(m_{Z})=0.118$, a SUSY threshold of 500 GeV and
$\tan\beta=5$.  The results depend weakly on these choices,
assuming $\tan\beta\approx2$-30.  Further, the Dirac masses and
mixings of the neutrinos and the mixings of the charged leptons
also get determined. We obtain:
\begin{equation}
m_{\nu_{\tau}}^{D}(M_{U})\,\approx\,100\mbox{-}120\,
\mbox{GeV};\,\,m_{\nu_{\mu}}^{D}(M_{U})\,\simeq\,8\,\mbox{GeV},
\label{e23}
\end{equation}

\begin{equation}
\theta_{\mu\tau}^{\ell}\,\approx\,-\,3\epsilon\,+\,\eta\,
\approx\,\sqrt{m_{\mu}/m_{\tau}}\,\left|\frac{-\,3\epsilon\,+
\,\eta}{3\epsilon\,+\,\eta}\right|^{1/2}\,\simeq\,0.437
\label{e24}
\end{equation}

\begin{equation}
m_{\nu_{e}}^{D}\,\simeq\,[\,9\epsilon^{'2}/(9\epsilon^{2}\,-
\,\sigma^{2})]\,m_{U}\,\simeq\,0.4\,\mbox{MeV}
\label{e25}
\end{equation}

\begin{equation}
\theta_{e\mu}^\ell\,\simeq\,\left|\frac{\eta'\,-\,3\epsilon'}
{\eta'\,+\,3\epsilon'}\right|^{1/2}\,\sqrt{m_{e}/m_{\mu}}\,
\simeq\,0.85\,\sqrt{m_{e}/m_{\mu}}\,\simeq\,0.06
\label{e26}
\end{equation}

\begin{equation}
\theta_{e\tau}^\ell\,\simeq\,\frac{1}{0.85}\,
\sqrt{m_{e}/m_{\tau}}\,(m_{\mu}/m_{\tau})\,\simeq\,0.0012\,.
\label{e27}
\end{equation}
In evaluating $\theta_{e\mu}^\ell$, we have
assumed $\epsilon'$ and $\eta'$ to be relatively positive.

Given the bizarre pattern of quark and lepton masses and mixings,
it seems remarkable that the simple and economical pattern of
fermion mass-matrices, motivated in part by the assumption of
flavor symmetries$^6$ which distinguish between the three families
and in large part by the group theory of G(224)/SO(10), gives an
overall fit to all of them [Eqs. (\ref{e18}) through (\ref{e22})]
which is good to within $10\%$. This includes the two successful
predictions on $m_{b}$ and $V_{cb}$ [Eqs.(\ref{e18}) and
(\ref{e19})].  Note that in supersymmetric unified theories, the
``observed'' value of $m_{b}(m_{b})$ and renormalization-group
studies suggest that, for a wide range of the parameter
$\tan\beta$, $m_{b}^{0}$ should in fact be about 10-20$\%$ {\it
lower} than $m_{\tau}^{0}$ \cite{Piercembmt}.  This is neatly
explained by the relation: $m_{b}^{0}\approx m_{\tau}^{0}(1 -
8\epsilon^{2})$ [Eq. (\ref{e18})], where exact equality holds in
the limit $\epsilon\rightarrow0$ (due to SU(4)-color), while the
decrease of $m^{0}_{b}$ compared to $m^{0}_{\tau}$ by
$8\epsilon^{2}\sim10\%$ is precisely because the off-diagonal
$\epsilon$-entry is proportional to B--L [see Eq. (\ref{e16})].

Specially intriguing is the result on $V_{cb}\approx0.045$ which
compares well with the observed value of $\simeq0.04$.  The
suppression of $V_{cb}$, compared to the value of $0.17 \pm 0.06$
obtained from Eq. (\ref{e13}), is now possible because the mass
matrices [Eq. (\ref{e16})] contain an antisymmetric component
$\propto\epsilon$.  That corrects the square-root formula
$\theta_{sb}=\sqrt{m_{s}/m_{b}}$ [appropriate for symmetric
matrices, see Eq.  (\ref{e11})] by the asymmetry factor
$|(\eta+\epsilon)/(\eta-\epsilon)|^{1/2}$ [see Eq. (19)], and
similarly for the angle $\theta_{ct}$.  This factor suppresses
$V_{cb}$ if $\eta$ and $\epsilon$ have opposite signs.  The
interesting point is that, {\it the same feature necessarily
enhances the corresponding mixing angle $\theta_{\mu\tau}^{\ell}$
in the leptonic sector}, since the asymmetry factor in this case
is given by $[(-3\epsilon+\eta)/(3\epsilon+\eta)]^{1/2}$ [see Eq.
(24)].  This enhancement of $\theta_{\mu\tau}^\ell$ helps to
account for the nearly maximal oscillation angle observed at
SuperK (as discussed below).  This intriguing correlation between
the mixing angles in the quark versus leptonic sectors---{\it that
is suppression of one implying enhancement of the other}---has
become possible only because of the $\epsilon$-contribution, which
is simultaneously antisymmetric and is proportional to B--L. That
in turn becomes possible because of the group-property of SO(10)
or a string-derived G(224)$^{7}$.

Taking stock, we see an impressive set of facts in favor of having
B--L as a gauge symmetry and in fact for the full
SU(4)-color-symmetry.  These include: (i) the suppression of
$V_{cb}$, together with the enhancement of
$\theta_{\mu\tau}^{\ell}$, mentioned above; (ii) the successful
relation $m_{b}^{0}\approx m_{\tau}^{0}(1-8\epsilon^{2})$; (iii)
the usefulness again of the SU(4)-color-relation $m(\nu_{\rm
Dirac}^{\tau})^{0}\approx m_{t}^{0}$ in accounting for
$m(\nu_{L}^{\tau})$ (see Section 4); (iv) the agreement of the
relation
$|m_{s}^{0}/m_{\mu}^{0}|=|(\epsilon^{2}-\eta^{2})/(9\epsilon^{2}-\eta^{2})|$
with the data, in that the ratio is naturally {\it less than} 1,
if $\eta\sim\epsilon$ [The presence of $9\epsilon^2$ in the
denominator is because the off-diagonal entry is proportional to
B--L.]; and finally (v), the need for (B--L)---as a local
symmetry, to implement baryogenesis via leptogenesis, as noted in
Section 1.

Turning to neutrino masses, while all the entries in the Dirac
mass matrix $N$ are now fixed, to obtain the parameters for the
light neutrinos, one needs to specify those of the Majorana mass
matrix of the RH neutrinos ($\nu^{e,\mu,\tau}_{R}$).  Guided by
economy and the assumption of hierarchy, we consider the following
pattern \cite{BabuWilczekPati}:
\begin{equation}
M_{\nu}^{R} = \left(\begin{array}{ccc}{x} & {0} & {z}
\\ {0} & {0} & {y} \\ {z} & {y} & {1}\end{array}\right)\,M_{R}\,.
\label{e28}
\end{equation}

As discussed in Section  \ref{Mass}, the magnitude of
$M_{R}\approx(5\mbox{-}10)\times10^{14}$ GeV can quite plausibly
be justified in the context of supersymmetric unification$^5$
[e.g. by using $M\approx M_{st}\approx4\times10^{17}$ GeV in Eq.
(\ref{e8})].  To the same extent, the magnitude of
$m(\nu_{\tau})\approx(1/10\mbox{-}1/30)$ eV, which is consistent
with the SuperK value, can also be anticipated by allowing for
$\nu_\mu-\nu_\tau$ mixing [see Ref. \cite{BabuWilczekPati}]. Thus
there are effectively three new parameters: $x$, $y$, and $z$.
Since there are six observables for the three light neutrinos, one
can expect three predictions. These may be taken to be
$\theta_{\nu_{\mu}\nu_{\tau}}^{\rm osc}$, $m_{\nu_{\tau}}$ [see
Eq. (\ref{e10})], and for example $\theta_{\nu_{e}\nu_{\mu}}^{\rm
osc}$.

Assuming successively hierarchical entries as for the Dirac mass
matrices, we presume that $|y|\sim1/10, |z|\leq|y|/10$ and
$|x|\leq z^{2}$. Now given that $m(\nu_{\tau})\sim1/20$ eV [as
estimated in Eq. (\ref{e10})], the MSW solution for the solar
neutrino puzzle \cite{MSW2} suggests that
$m(\nu_{\mu})/m(\nu_{\tau})\approx1/8\mbox{-}1/20$. With
hierarchical neutrino masses, the higher value of the mass-ratio
(like 1/8) holds only for the large angle MSW solution (see
below). With the mass-ratio being in the range of 1/8-1/20, one
obtains: $|y|\approx(1/17\mbox{\,\,to\,\,}1/21)$, with $y$ having
the same sign as $\epsilon$ [see Eq. (\ref{e17})]. This solution
for $y$ obtains only by assuming that $y$ has a hierarchical value
$O(1/10)$ rather than $O(1)$. Combining now with the mixing in the
$\mu$-$\tau$ sector determined above [see Eq. (\ref{e24})], one
can then determine the $\nu_{\mu}\mbox{-}\nu_{\tau}$ oscillation
angle. The two predictions of the model for the neutrino-system
are then:
\begin{equation}
m(\nu_{\tau})\,\approx\,(1/10\,\mbox{-}\,1/30)\,\mbox{eV}
\label{e29n}
\end{equation}
\begin{equation}
\theta_{\nu_{\mu}\nu_{\tau}}^{\rm
osc}\,\simeq\,\theta_{\mu\tau}^{\ell} -
 \theta_{\mu\tau}^{\nu}\,\simeq
\,\left(0.437\,+\,\sqrt{\frac{m_{\nu_{2}}}{m_{\nu_{3}}}}\,\right)\, .
\label{e29}
\end{equation}
Thus,
\begin{equation}
\label{eq31} {\sin}^{2}\,2\theta_{\nu_{\mu}\nu_{\tau}}^{\rm osc}=
(0.99,0.975,0.92,0.87)
\end{equation}
for
\begin{equation}
m_{\nu_{2}}/m_{\nu_{3}}=(1/8,1/10,1/15,1/20)\,. \label{e30}
\end{equation}
Both of these predictions are extremely successful.\footnote{In
writing Eq. (\ref{eq31}), the small angle approximation exhibited
in Eq. (\ref{e29}) is replaced by the more precise expression,
given in Eq. (12) of Ref. \cite{BabuWilczekPati}, with the further
understanding that $\sqrt{m_\mu/m_\tau}$ appearing in Eq. (12) (of
Ref. \cite{BabuWilczekPati})
 is replaced by the $\mu$-$\tau$ mixing angle $\approx$ 0.437.}

Note the interesting point that the MSW solution, and the
requirement that $|y|$ should have a natural hierarchical value
(as mentioned above), lead to $y$ having the same sign as
$\epsilon$.  Now, that (it turns out) implies that the two
contributions in Eq.  (\ref{e29}) must {\it add} rather than
subtract, leading to an {\it almost maximal oscillation angle\,}
\cite{BabuWilczekPati}.  The other factor contributing to the
enhancement of $\theta_{\nu_{\mu}\nu_{\tau}}^{\rm osc}$ is, of
course, also the asymmetry-ratio which increases
$|\theta_{\mu\tau}^{\ell}|$ from 0.25 to 0.437 [see Eq.
(\ref{e24})]. We see that one can derive rather plausibly a large
$\nu_{\mu}\mbox{-}\nu_{\tau}$ oscillation angle
$\sin^{2}\,2\theta_{\nu_{\mu}\nu_{\tau}}^{\rm osc}\geq0.92$,
together with an understanding of hierarchical masses and mixings
of the quarks and the charged leptons, while maintaining a large
hierarchy in the seesaw derived neutrino masses
($m_{\nu_{2}}/m_{\nu_{3}}=1/8\mbox{-}1/15$), all within a unified
framework including both quarks and leptons.  In the example
exhibited here, the mixing angles for the mass eigenstates of
neither the neutrinos nor the charged leptons are really large, in
that $\theta_{\mu\tau}^{\ell}\simeq0.437\simeq23^{\circ}$ and
$\theta_{\mu\tau}^{\nu}\simeq(0.22\mbox{-}0.35)
\approx(13\mbox{-}20.5)^{\circ}$, {\it yet the oscillation angle
obtained by combining the two is near-maximal.} This contrasts
with most works in the literature in which a large oscillation
angle is obtained either entirely from the neutrino sector (with
nearly degenerate neutrinos) or almost entirely from the charged
lepton sector.

\subsection*{Small Versus Large Angle MSW Solutions}
\label{MSWSolutions}

In considerations of $\nu_e$-$\nu_\mu$ and $\nu_e$-$\nu_\tau$
oscillation angles, tiny {\it intrinsic} non-diagonal Majorana
masses $\sim 10^{-3}$ eV of the LH neutrinos leading to
$\nu^e_L\nu^\mu_L$ and $\nu^e_L\nu^\tau_L$-mixings, which can far
exceed those induced by the standard see-saw mechanism, can be
rather important, especially for $\nu_e$-$\nu_\mu$ mixing. As
explained below, such intrinsic masses can arise quite naturally
through higher dimensional operators and can lead to the large
angle MSW solution of the solar neutrino puzzle.

Let us first ignore the intrinsic Majorana masses of the LH
neutrinos and include only those that arise through the standard
see-saw mechanism, involving the superheavy Majorana masses of the
RH neutrinos, with a pattern given, for example, by Eq.
(\ref{e28}). Note that, while
$M_{R}\approx(5\mbox{-}15)\times10^{14}$ GeV and $y\approx-1/20$
are better determined, the parameters $x$ and $z$ can not be
obtained reliably at present because very little is known about
observables involving $\nu_{e}$. Taking, for concreteness,
$m_{\nu_{e}}\approx(10^{-5}\mbox{-}10^{-4})$ (1 to few)) eV and
$\theta^{\rm
osc}_{e\tau}\approx\theta^{\ell}_{e\tau}-\theta^{\nu}_{e\tau}
\approx10^{-3}\pm0.03$ as inputs, we obtain:
$z\sim(1$-$5)\times10^{-3}$ and $x\sim($1 to
few)$(10^{-6}\mbox{-}10^{-5})$, in accord with the guidelines of
$|z|\sim|y|/10$ and $|x|\sim z^{2}$. This in turn yields:
$\theta^{\rm
osc}_{e\mu}\approx\theta^{\ell}_{e\mu}-\theta^{\nu}_{e\mu}
\approx0.06\pm0.015$. Note that the mass of
$m_{\nu_{\mu}}\sim3\times10^{-3}$ eV, that follows from a natural
hierarchical value for $y\sim-(1/20)$, and $\theta_{e\mu}$ as
above, go well with the small angle MSW explanation of the solar
neutrino puzzle. In short the framework presented so far, that
neglects intrinsic Majorana masses of the LH neutrinos altogether,
generically tends to yield the small angle MSW solution.

As alluded to above, we now observe that small intrinsic
non-seesaw masses of the LH neutrinos $\sim 10^{-3}$ eV, which
could mix $\nu_{eL}$ and $\nu_{\mu L}$, can, however, arise quite
naturally through higher dimensional operators in the
superpotential of the form \footnote{Such a term can be induced in
the presence of, for example, a singlet S and a ten-plet (denoted
by $\hat\mathbf{10}$), both having GUT-scale masses, and
possessing renormalizable couplings of the form $a_i\mathbf{16}_i
\mathbf{16}_H \widehat\mathbf{10}$,
$b\widehat\mathbf{10}\,\mathbf{10}_HS$, $M_SSS$ and
$\widehat{M}\widehat\mathbf{10}^2$. In this case, $\kappa
_{12}/M^3_{\rm GUT}=a_1a_2b/ (\widehat{M}^2M_S)$.}: $W\supset
\kappa_{12}\mathbf{16}_1\mathbf{16}_2\mathbf{16}_H \mathbf{16}_H
\mathbf{10}_H\mathbf{10}_H/M^3_{\rm GUT}$. One can verify that
such a term would lead to an {\it intrinsic} Majorana mixing mass
term of the form $m_{12}^{(0)}\nu^e_L\nu^\mu_L$, with a strength
given by
$m_{12}^{(0)}\approx\kappa_{12}(\langle\mathbf{16}_H\rangle/M_{\rm
GUT})^2 (175\mbox{ GeV})^2/M_{\rm GUT}\approx(1.5\mbox{-}6)\times
10^{-3}eV$, where we have put
$\langle\mathbf{16}_H\rangle\approx(1\mbox{-}2)M_{\rm GUT}$ and
$M_{\rm GUT}\approx 2\times 10^{16}$ GeV. Such an intrinsic
Majorana mixing mass $\sim 10^{-3}$ eV, though small, is still
much larger than what one would get for the corresponding term
from the standard see-saw mechanism. Now, as discussed above, the
diagonal $(\nu_L^\mu\nu_L^\mu)$ mass-term, arising from the
standard see-saw mechanism can naturally be of order (3-8)$\times
10^{-3}$ eV (for $|y|\approx 1/20$ to 1/15, say). In addition, the
intrinsic contribution of the type mentioned above may in general
also contribute to the diagonal
 $(\nu_L^\mu\nu_L^\mu)$ mass (depending upon flavor symmetries) which can be
 (few)$\times 10^{-3}$ eV. Thus, taking the net values of $m_{22}\approx
 (6\mbox{-}7)\times 10^{-3}$ eV (say), $m^{(0)}_{12}\approx
 (3\mbox{-}4)\times 10^{-3}$ eV, and $m^{(0)}_{11}\lsim
 (1\mbox{-}2)\times 10^{-3}$ eV, which are all very plausible, we obtain
 $m_{\nu_\mu}\approx (6\mbox{-}7)\times 10^{-3}$ eV, $m_{\nu_e}\sim 1
 \times 10^{-3}$ eV, so that $\Delta m^2_{12}\approx (3.6\mbox{-}5)
 \times 10^{-5}$ eV$^2$, and $\sin^22\theta_{12}^{\rm osc}\approx 0.6\mbox{-}0.7$.
 This goes well with the large angle MSW solution of the solar neutrino puzzle,
 which is now favored over the small angle solution by the SuperK data \cite{newref}.

In summary, the intrinsic non-seesaw contribution to the Majorana
masses of the LH neutrinos quite plausibly has the right magnitude
for $\nu_e$-$\nu_\mu$ mixing, so as to lead to the rather large
oscillation angle as mentioned above, in accord with the data.  In
contrast to the case of the $\nu_\mu$-$\nu_\tau$ oscillation
angle, however, given the smallness of the entries involving the
first two families, the relatively large angle solution for
$\nu_e-\nu^\mu$ oscillation may not be regarded as a firm
prediction of the SO(10)/G(224)-framework presented here.  It is
nevertheless a very reasonable possibility.

It is worth noting that although the superheavy Majorana masses of
the RH neutrinos cannot be observed directly, they can be of
cosmological significance.  The pattern given above and the
arguments given in Section \ref{Need} and in this section suggests
that $M(\nu_{R}^{\tau})\approx(5\mbox{-}15)\times10^{14}$ GeV,
$M(\nu_{R}^{\mu})\approx(1\mbox{-}4)\times10^{12}$ GeV (for
$|y|\approx1/20$); and
$M(\nu_{R}^{e})\sim(1/2\mbox{-}10)\times10^9$ GeV (for
$x\sim(1/2\mbox{-}10)10^{-6}>z^2$).  A mass of
$\nu_{R}^{e}\sim10^{9}$ GeV is of the right magnitude for
producing $\nu_{R}^{e}$ following reheating and inducing lepton
asymmetry in $\nu_{R}^{e}$ decay into $H^{0}+\nu_{L}^{i}$, that is
subsequently converted into baryon asymmetry by the electroweak
sphalerons \cite{KuzminRubakov,LeptoB}.

In summary, we have proposed an economical and predictive pattern
for the Dirac mass matrices, within the SO(10)/G(224)-framework,
which is remarkably successful in describing the observed masses
and mixings of {\it all} the quarks and charged leptons.  It leads
to five predictions for just the quark- system, all of which agree
with observation to within 10\%.  The same pattern, supplemented
with a similar structure for the Majorana mass matrix, accounts
for both the nearly-maximal $\nu_{\mu}$-$\nu_{\tau}$ oscillation
angle and a (mass)$^2$-difference $\Delta m^2(\nu_\mu\nu_\tau)
\sim$ (1/20 eV)$^2$, suggested by the SuperK data. Given this
degree of success, it makes good sense to study proton decay
concretely within this SO(10)/G(224)-framework. The results of
this study \cite{BabuWilczekPati,JCP_Erice} are presented in the
next section, together with an update.

Before turning to proton decay, it is worth noting that much of our
discussion of fermion masses and mixings, including those of the
neutrinos, is essentially unaltered if we go to the limit
$\epsilon'\rightarrow0$ of Eq.  (28).  This limit clearly involves:
\[
m_{u}\,=\,0\,,\,\,\,\,\theta_{C}\,\simeq\,\sqrt{m_{d}/m_{s}}
\,,\,\,\,\,m_{\nu_{e}}\,=\,0\,,\,\,\,\,\theta_{e\mu}^{\nu}\,
=\,\theta_{e\tau}^{\nu}\,=\,0\,   \]
\begin{equation}
|V_{ub}|\,\simeq\,\sqrt{\frac{\eta\,-\,\epsilon}{\eta\,+\,
\epsilon}}\,\sqrt{m_{d}/m_{b}}\,(m_{s}/m_{b})\,\simeq\,(2.1)
(0.039)(0.023)\,\simeq\,0.0019 \ .\label{e32}
\end{equation}
All other predictions remain unaltered.  Now, among the observed
quantities in the list above, $\theta_C\simeq\sqrt{m_{d}/m_{s}}$
is a good result.  Considering that $m_{u}/m_{t}\approx10^{-5}$,
$m_{u}=0$ is also a pretty good result. There are of course
plausible small corrections which could arise through Planck scale
physics; these could induce a small value for $m_{u}$ through the
(1,1)-entry $\delta\approx10^{-5}$.  For considerations of proton
decay, it is worth distinguishing between these two {\it extreme}
variants which we will refer to as cases I and II respectively.
\begin{eqnarray}
\mbox{Case I
:}&&\epsilon'\,\approx\,2\,\times\,10^{-4}\,,\,\,\,\delta\,=\,0
\nonumber \\[1em]
\mbox{Case II:}&&\delta\,\approx\,10^{-5}\,,\,\,\,\epsilon'\,=
0\,. \label{e33}
\end{eqnarray}
It is worth noting that the observed value of $|V_{ub}|\approx
0.003$ favors a non-zero value of $\epsilon'$ $(\approx
(1$-$2)\times 10^{-4})$. Thus, in reality, $\epsilon'$ may not be
zero, but it may lie in between the two extreme values listed
above. In this case, the predicted proton lifetime for the
standard $d=5$ operators would be intermediate between those for
the two cases, presented in Section \ref{Expectations}.

\section{\large Expectations for Proton Decay in Supersymmetric
Unified \hfill \break Theories} \label{Expectations}

\subsection{Preliminaries}\label{Preliminaries}

Turning to the main purpose of this talk, I present now the reason why
the unification framework based on SUSY SO(10) or G(224), together with
the understanding of fermion masses and mixings discussed above,
strongly suggest that proton decay should be imminent.

Recall that supersymmetric unified theories (GUTs) introduce two
new features to proton decay:  (i) First, by raising $M_{X}$ to a
higher value of about $2\times10^{16}$ GeV (contrast with the
non-supersymmetric case of nearly $3\times 10^{14}$ GeV), they
strongly suppress the gauge-boson-mediated $d=6$ proton decay
operators, for which $e^{+}\pi^{0}$ would have been the dominant
mode (for this case, one typically obtains:
$\Gamma^{-1}(p\rightarrow e^{+}\pi^{0})|_{d=6}\approx10^{35\pm1}$
years).  (ii) Second, they generate $d=5$ proton decay operators
\cite{Sakai} of the form $Q_{i}Q_{j}Q_{k}Q_{l}/M$ in the
superpotential, through the exchange of color triplet Higginos,
which are the GUT partners of the standard Higgs(ino) doublets,
such as those in the $\mathbf{5}+\overline\mathbf{5}$ of SU(5) or
the 10 of SO(10). Assuming that a suitable doublet-triplet
splitting mechanism provides heavy GUT-scale masses to these color
triplets and at the same time light masses to the doublets (see
e.g, the Appendix), these ``standard" $d=5$ operators, suppressed
by just one power of the heavy mass and the small Yukawa
couplings, are found to provide the dominant mechanism for proton
decay in supersymmetric GUT
\cite{PierceBabuKolda,Ellis,MSW,Hisano,BabuBarr}.

Now, owing to (a) Bose symmetry of the superfields in $QQQL/M$,
(b) color antisymmetry, and especially (c) the hierarchical Yukawa
couplings of the Higgs doublets, it turns out that these standard
$d=5$ operators lead to dominant $\overline{\nu}K^{+}$ and
comparable $\overline{\nu}\pi^{+}$ modes, but in all cases to
highly suppressed $e^{+}\pi^{0}$, $e^{+}K^{0}$ and even
$\mu^{+}K^{0}$ modes.  For instance, for minimal SUSY SU(5), one
obtains (with $\tan\beta\leq20$, say):
\begin{equation}
[\,\Gamma(\mu^{+}K^{0})/\Gamma(\overline{\nu}K^{+})\,]^{SU(5)}_{std}
\,\sim\,[m_{u}/(m_{c}\,\sin^2\theta_c)]^2\,R\,\approx\,10^{-3}\,,
\label{e34}
\end{equation}
where $R\approx0.1$ is the ratio of the
relevant $|$matrix element$|^{2}\times$(phase space), for the two
modes.

It was recently pointed out that in SUSY unified theories based on
SO(10) or G(224), which assign heavy Majorana masses to the RH
neutrinos, there exists a new set of color triplets and thereby
very likely a {\it new source} of $d=5$ proton decay operators
\cite{BPW1}. For instance, in the context of the minimal set of
Higgs multiplets \footnote{The origin of the new $d=5$ operators
in the context of other Higgs multiplets, in particular in the
cases where $\mathbf{126}_{H}$ and $\overline\mathbf{126}_{H}$ are
used to break B--L, has been discussed in Ref. \cite{BPW1}.}
$\{\mathbf{45}_{H},\mathbf{16}_{H},\overline\mathbf{16}_{H}$ and
$\mathbf{10}_{H}\}$ (see Section \ref{Understanding}), these new
$d=5$ operators arise by combining three effective couplings
introduced before:---i.e., (a) the couplings
$f_{ij}\mathbf{16}_{i}\mathbf{16}_{j}
\overline{\mathbf{16}}_{H}\overline{\mathbf{16}}_{H}/M$ [see Eq.
(\ref{e7})] that are required to assign Majorana masses to the RH
neutrinos, (b) the couplings
$g_{ij}\mathbf{16}_{i}\mathbf{16}_{j}\mathbf{16}_{H}
\mathbf{16}_{H}/M$, which are needed to generate non-trivial CKM
mixings [see Eq. (\ref{e15})], and (c) the mass term
$M_{16}\mathbf{16}_{H}\overline{\mathbf{16}}_{H}$.  For the
$f_{ij}$ couplings, there are two possible SO(10)-contractions
(leading to a 45 or a 1) for the pair
$\mathbf{16}_i\overline{\mathbf{16}}_H$, both of which contribute
to the Majorana masses of the RH neutrinos, but only the
non-singlet contraction (leading to $\mathbf{45}$), would
contribute to $d=5$ proton decay operator.  In the presence of
non-perturbative quantum gravity, one would in general expect the
two contractions to have comparable strength.  Furthermore, the
couplings of $\mathbf{45's}$ lying in the string-tower or possibly
below the string-scale, and likewise of singlets, to the
$\mathbf{16}_i\cdot\overline{\mathbf{16}}_H$-pair, would
respectively generate the two contractions.  It thus seems most
likely that both contractions would be present, having comparable
strength. Allowing for a difference between the relevant
projection factors for $\nu_R$ masses versus proton decay, and
also for the fact that both contractions contribute to the former,
but only the non-singlet one (i.e. 45) to the latter, we would set
the relevant $f_{ij}$ coupling for proton decay to be
$(f_{ij})_p\equiv(f_{ij})_\nu\cdot K$, where $(f_{ij})_\nu$
defined in Section \ref{Mass} directly yields $\nu_R$ - masses
[see Eq. (\ref{e8})]; and K is a relative factor, which
generically is expected to be of order unity. \footnote{For the
special case of $K=0$ (which would arise if only the
singlet-contraction of
$(\mathbf{16}_i\cdot\overline{\mathbf{16}}_H)$ contributes), the
new $d=5$ operators shown in Eq.  (\ref{e35}) would not, of
course, contribute to proton decay.}  As a plausible range, we
will take $K\approx 1/5$ to 2 (say).  In the presence of the
non-singlet contraction, the color-triplet Higginos in
$\overline{\mathbf{16}}_{H}$ and $\mathbf{16}_{H}$ of mass
$M_{16}$ can be exchanged between $\tilde{q}_{i}q_{j}$ and
$\tilde{q}_{k}q_{l}$-pairs (correspondingly, for G(224), the color
triplets would arise from $(1,2,4)_H$ and $(1, 2,
\overline{4})_H$).  This exchange generates a new set of $d=5$
operators in the superpotential of the form
\begin{equation}
W_{\rm
new}\,\propto\,(f_{ij})_\nu\,g_{kl}K\,(\mathbf{16}_{i}\,\mathbf{16}_{j})\,
(\mathbf{16}_{k}\,\mathbf{16}_{l})\,\langle\overline{\mathbf{16}}_{H}\rangle\,
\langle\mathbf{16}_{H}\rangle/M^2\,\times (1/M_{16}), \label{e35}
\end{equation}
which induce proton decay.  Note that these operators
depend, through the couplings $f_{ij}$ and $g_{kl}$, both on the
Majorana and on the Dirac masses of the respective fermions.  {\it This
is why within SUSY SO(10) or G(224), if the generic case of $K\neq 0$
holds, proton decay gets intimately linked to the masses and mixings of
all fermions, including neutrinos.}

\subsection{Framework for Calculating Proton Decay Rate}
\label{Framework}

To establish notations, consider the case of minimal SUSY SU(5)
and, as an example, the process
$\tilde{c}\tilde{d}\rightarrow{\bar{s}}{\bar{\nu}_{\mu}}$, which
induces $p\rightarrow\overline{\nu}_{\mu}K^{+}$. Let the strength
of the corresponding $d=5$ operator, multiplied by the product of
the CKM mixing elements entering into wino-exchange vertices,
(which in this case is $\sin\theta_{C}\cos\theta_{C})$ be denoted
by $\widehat{A}$. Thus (putting $\cos\theta_{C} =1$), one obtains:
\begin{eqnarray}
\widehat{A}_{\tilde{c}\tilde{d}}(SU(5))
 & = &(h_{22}^{u}\,h_{12}^{d}/M_{H_{C}})\,\sin\theta_{c}\nonumber
 \\ & \simeq &(m_{c}m_{s}\,\sin^{2}\theta_{C}/v_{u}^{2})
\,(\tan\beta/M_{H_{C}})\nonumber\\
& \simeq & (1.9\times10^{-8})\,(\tan\beta/M_{H_C})\\
 & \approx & (2\times10^{-24}\,\mbox{GeV$^{-1}$})
\,(\tan\beta/2)\,(2\times10^{16}\,\mbox{GeV}/M_{H_{C}})\nonumber,
\label{e36}
\end{eqnarray}
where $\tan\beta\equiv{v}_{u}/v_{d}$, and we have put $v_{u}=174$
GeV and the fermion masses extrapolated to the
unification-scale---i.e. $m_{c}\simeq300$ MeV and $m_{s}\simeq40$
MeV.  The amplitude for the associated four-fermion process
$dus\rightarrow\overline{\nu}_{\mu}$ is given by:
\begin{equation}
A_5(dus\,\rightarrow\,\overline{\nu}_{\mu})\,=\,
\widehat{A}_{\tilde{c}\tilde{d}}\,\times\,(2f) \label{e37}
\end{equation}
where $f$ is the loop-factor associated with wino-dressing.
Assuming $m_{\tilde{w}}\ll{m}_{\tilde{q}}\sim{m}_{\tilde{l}}$, one
gets: $f\simeq(m_{\tilde{w}}/m^{2}_{\tilde{q}})(\alpha_{2}/4\pi)$.
Using the amplitude for $(du)(s\nu_\ell$), as in Eq. (\ref{e37}),
($\ell=\mu$ or $\tau$), and the recently obtained matrix element
and renormalization effects (see below), one then obtains
\cite{MSW,Hisano,BabuBarr,BabuWilczekPati,JCP_Erice}:
\begin{eqnarray}
\Gamma^{-1}(p\,\rightarrow\,\overline{\nu}_{\tau}K^+)\,\approx\,(0.15
\times10^{31})\,\mbox{years}\times(0.32/A_L)^2 \\[1ex]
&&\hspace{-7cm}\times
\left(\frac{0.93}{{A}_{S}}\right)^2\,\left[\frac{0.014\,\mbox{GeV}^3}
{\beta_{H}}\right]^{2}\,\left[\frac{(1/6)}{(m_{\tilde{W}}/m_{\tilde{q}})}
\right]^{2} \left[\frac{m_{\tilde{q}}}{1.2\,\mbox{TeV}}\right]^2\,
\left[\frac{2\times10^{-24}\,\mbox{GeV}^{-1}}{\widehat{A}(\overline{\nu})}\right]
^2\,. \nonumber\label{e38}
\end{eqnarray}
Here $\beta_{H}$ denotes the hadronic matrix element defined by
$\beta_H u_L(\vec k)\equiv  \epsilon_{\alpha\beta\gamma}
\VEV{0|(d_L^\alpha u_L^\beta)u_{L}^{\gamma}|p,\vec k}$. While the
range $\beta_H=(0.003\mbox{-}0.03)$ GeV$^3$ has been used in the
past \cite{Hisano}, given that one  lattice calculation yields
$\beta_H=(5.6\pm0.5)\times10^{-3}$ GeV$^3$ \cite{JLQCD99}, and a
recent improved calculation yields $\beta_H\approx 0.014$ GeV$^3$
\cite{Aoki} (whose systematic errors that may arise from scaling
violations and quenching are hard to estimate \cite{Aoki}), we
will take as a conservative, but plausible, range for $\beta_H$ to
be given by $(0.014$ GeV$^3)(1/2 - 2)$. (Compare this with the
range for $\beta_H = (0.006$ GeV$^3)(1/2 - 2)$ as used in Ref.
\cite{BabuWilczekPati}). $A_S$ denotes the short-distance
renormalization effect for the $d=5$ operator which arises owing
to extrapolation between the GUT and the SUSY-breaking scales
\cite{Ellis,Hisano,Turzynski}. The average value of $A_S=0.67$,
given in Ref. \cite{Hisano} for $m_t=100$ GeV, has been used in
most early estimates. For $m_t=175$ GeV, one would,
 however, have  $A_S\approx 0.93$ to 1.2
 \cite{Turzynski}. Conservatively, I would use $A_S=0.93$; this would enhance
 the rate by a factor of two compared with previous estimates. $A_L$ denotes
 the long-distance renormalization effect of the $d=6$ operator due to QCD
 interaction that arises due to extrapolation between the SUSY breaking scale
 and 1 GeV \cite{Ellis}. Using the two-loop expression for $A_L$
 \cite{Arafune}, together with the two-loop value for $\alpha_3$, Babu and I
 find: $A_L\approx 0.32$, in contrast to $A_L\approx 0.22$, used in previous
 works\footnote{In most previous works starting with Ref. \cite{Ellis}
 through \cite{BabuBarr}, as well as in Refs. \cite{BabuWilczekPati}
 and \cite{JCP_Erice}, the one-loop value of $A_L$ was taken to be 0.22.
 It was, however, noted in Refs. \cite{Arafune} and \cite{Dermisek} that
 there is a numerical error in the evaluation of the one-loop expression for
 $A_L$ \cite{Ellis}, and that the correct value for
 $A_L({\rm one-loop})\approx 0.43$ (this remained unnoticed by most authors).
 The two-loop value for $A_L$ (as stated above) is nearly 0.32, which is
 lower than 0.43 but higher that the previously used value of 0.22.}.
 In what follows, I would use $A_L\approx 0.32$.
 This by itself would also increase the rate by a
 factor of (0.32/0.22)$^2\approx$2, compared to the previous estimates
\cite{Ellis,MSW,Hisano,BabuBarr,BabuWilczekPati,JCP_Erice}.
Including the enhancements in both $A_S$ and $A_L$, we thus see
that the net increase in the proton decay rate solely due to new
evaluation of renormalization effects is nearly a factor of four,
compared to the previous estimates (including that in Ref.
\cite{BabuWilczekPati}).

Note that the familiar factors that appear in the expression for
proton lifetime---i.e., $M_{H_{C}}$, ($1+y_{tc}$) representing the
interference between the $\tilde{t}$ and $\tilde{c}$
contributions, and $\tan\beta$ (see e.g.  Ref. \cite{Hisano} and
discussion in the Appendix of Ref. \cite{BabuWilczekPati})---are
all effectively contained in $\widehat{A}(\overline{\nu})$.  In
Ref. \cite{BabuWilczekPati}, guided by the demand of naturalness
(i.e. absence of excessive fine tuning) in obtaining the Higgs
boson mass, squark masses were assumed to lie in the range of 1
TeV$(1/\sqrt{2} - \sqrt{2})$, so that $m_{\tilde{q}}\lsim 1.4$TeV.
Recent work, based on the notion of focus point supersymmetry
however suggests that squarks may be considerably heavier without
conflicting with the demands of naturalness \cite{Feng+}.  In the
interest of obtaining a conservative upper limit on proton
lifetime, we will therefore allow squark masses to be as heavy as
about 2.4 TeV and as light as perhaps 600 GeV. \footnote{ We
remark that if the recently reported (g-2)-anomaly for the muon
\cite{g2}, together with reevaluation of the contribution from
light by light-scattering \cite{lightbylight}, is attributed to
supersymmetry \cite{SUSYg2}, one would need to have extremely
light s-fermions [i.e. $m_{\tilde{l}}\approx$ 200 - 400 GeV (say)
and correspondingly, for promising mechanisms of SUSY-breaking,
$m_{\tilde{q}}\lsim 300 - 600 $ GeV (say)], and simultaneously
relatively large $\tan\beta (\approx 6$-$24)$. However, not
worrying about grand unification, such light s-fermions, together
with large or very large $\tan\beta$ would typically be in gross
conflict with the limits on the edm's of the neutron and the
electron, unless one can explain naturally the occurrence of
minuscule phases $(\lsim 1/200$ to $1/500$) and/or large
cancellation.  Thus, if the $(g-2)_{\mu}$-anomaly turns out to be
real, it may well find a non-supersymmetric explanation, in accord
with the edm-constraints which ordinarily seem to suggest that
squarks are (at least) moderately heavy ($m_{\tilde{q}}\gsim 0.6 -
1$ TeV, say), and $\tan\beta$ is not too large ($\lsim 3$ to $10$,
say).  We mention in passing that the extra vector---like
matter---specially a $16 + \overline{16}$ of SO(10)---as proposed
in the so-called extended supersymmetric standard model (ESSM)
\cite{BabuJi,BabuPatiStrem}, with the heavy lepton mass being of
order 200 GeV, can provide such an explanation \cite{BabuJCP2}.
Motivations for the case of ESSM, based on the need for (a)
removing the mismatch between MSSM and string unification scales,
and (b) dilaton-stabilization, have been noted in Ref.
\cite{BabuJi}.  Since ESSM is an interesting and viable variant of
MSSM, and would have important implications for proton decay, we
will present the results for expected proton decay rates for the
cases of both MSSM and ESSM in the discussion to follow.}

Allowing for plausible and rather generous uncertainties in the matrix
element and the spectrum we take:
\[
\beta_{H}\,=\,(0.014\,\mbox{GeV}^3)\,(1/2\,\mbox{-}\,2)
\]
\begin{equation}
m_{\tilde{w}}/m_{\tilde{q}}\,=\,1/6\,(1/2\,\mbox{-}\,2)\,,\,\,\,\,
{\rm and}\,\,\,\,m_{\tilde{q}}\,\approx\,m_{\tilde{\ell}}\,\approx\,1.2\,
\mbox{TeV}\,(1/2\,\mbox{-}\,2)\,.
\label{e39}
\end{equation}
Using Eqs. (39--40), we get:
\begin{equation}
\Gamma^{-1}(p\,\rightarrow\,\overline{\nu}_{\tau}K^{+}) \approx
(0.15\times10^{31}\,\mbox{years})\,[\,2\times10^{-24}\,
\mbox{GeV}^{-1}/\widehat{A}(\overline{\nu}_\ell)\,]^{2} \times
\{64\, \mbox{-}\,1/64\,\}\,. \label{e40}
\end{equation}
Note that the curly bracket would acquire its upper-end value of
64, which would serve towards maximizing proton lifetime, only
provided all the uncertainties in Eq. (\ref{e40}) are stretched to
the extreme so that $\beta_H=0.007$ GeV$^3$,
$m_{\tilde{W}}/m_{\tilde{q}}\approx 1/12$ and
$m_{\tilde{q}}\approx 2.4$ TeV. This relation, as well as Eq. (39)
are general, depending only on
$\widehat{A}(\overline{\nu}_{\ell})$ and on the range of
parameters given in  Eq. (40). They can thus be used for both
SU(5) and SO(10).

The experimental lower limit on the inverse rate for the
$\bar{\nu}K^{+}$ modes is given by Ref. \cite{SKlimit},
\begin{equation}
\left[\sum_{\ell}\,\Gamma(p\,\rightarrow\,\overline{\nu}_{\ell}
K^{+})\right]^{-1}_{\rm expt} \geq 1.9\times
10^{33}\,\mbox{years}\,. \label{e41}
\end{equation}
Allowing for all the uncertainties to stretch in the same
direction (in this case, the curly bracket = 64), and assuming
that just one neutrino flavor (e.g. $\nu_{\mu}$ for SU(5))
dominates, the   observed limit (Eq. (\ref{e41})) provides an
upper bound on the amplitude\footnote{If there are sub-dominant
$\overline{\nu}_{i}K^{+}$ modes with branching ratio $R$, the
right side of Eq. (\ref{e42}) should be divided by $\sqrt{1+R}$.}:
\begin{equation}
\widehat{A}(\overline{\nu}_{\ell})\,\leq\,0.46\times10^{-24}\,\mbox{GeV}^{-1}
\label{e42}
\end{equation}
which holds for both SU(5) and SO(10).  Recent theoretical
analyses based on LEP-limit on Higgs mass $(\gsim 114$ GeV),
together with certain assumptions about MSSM parameters (as in
CMSSM) and/or constraint from muon g-2 anomaly \cite{g2} suggest
that $\tan\beta\gsim$ 3 to 5 \cite{Arnowitt}. In the interest of
getting a conservative upper limit on proton lifetime, we will
therefore use, as a conservative lower limit, $\tan\beta\geq 3$.
We will however exhibit relevant results often as a function of
$\tan\beta$ and exhibit proton lifetimes corresponding to higher
values of $\tan\beta$ as well. For minimal SU(5), using Eqs. (37)
and (\ref{e42}) and, conservatively $\tan\beta\geq 3$, one obtains
a lower limit on $M_{HC}$  given by:
\begin{equation}
M_{HC}\,\geq\,13\times10^{16}\,\mbox{GeV}\,\,\,(\mbox{SUSY
SU}(5))\ . \label{e43}
\end{equation}
At the same time, gauge coupling unification in SUSY SU(5) strongly suggests
$M_{HC}\leq (1/2\mbox{-}1)\times 10^{16}$ GeV. (See Ref. \cite{Murayama} where
an even more stringent upper bound on $M_{HC}$ is suggested.)
 Thus we already see a
conflict, in the case of minimal SUSY SU(5), between the
experimental limit on proton lifetime on the one hand, and
coupling unification and constraint on $\tan\beta$ on the other
hand. To see this conflict another way, if we keep $M_{HC}\leq
10^{16}$ GeV (for the sake of coupling unification) we  obtain
from Eq. (37): $\widehat{A}(\mbox{SU}(5))\geq
5.7\times10^{-24}\,\mbox{GeV}^{-1}(\tan\beta/3)$. Using Eq.
(\ref{e40}), this in turn implies that
\begin{equation}
\Gamma^{-1}(p\,\rightarrow\,\overline{\nu}K^+)\,\leq\,
   1.2\times10^{31}\,\mbox{years}
   \times(3/\tan\beta)^2\,\,\,\,\,(\mbox{SUSY\ SU}(5))\ .
\label{e44}
\end{equation}
For $\tan\beta \geq 3$, a lifetime of $1.2\times 10^{31}$ years is
thus a most conservative upper limit. In practice, it is unlikely
that all the uncertainties, including these in $M_{HC}$ and
$\tan\beta$, would stretch in the same direction to nearly extreme
values so as to prolong proton lifetime.  Given the experimental
lower limit [Eq. (\ref{e41})], we  see that minimal SUSY SU(5) is
already excluded by a large margin by  proton decay-searches. This
is in full accord with the conclusion reached by other authors
(see especially Ref. \cite{Murayama}).
 We have of course noted in
Section \ref{Mass} that SUSY  SU(5) does not go well with
neutrino oscillations observed at  SuperK.

Now, to discuss proton decay in the context of supersymmetric
SO(10), it is necessary to discuss first the mechanism for
doublet-triplet splitting.  Details of this discussion may be
found in Ref. \cite{BabuWilczekPati}.  A synopsis is presented in
the Appendix.

\subsection{Proton Decay in Supersymmetric SO(10)}

The calculation of the amplitudes $\widehat{A}_{\rm std}$ and
$\widehat{A}_{\rm new}$ for the standard  and the new operators
for the SO(10) model, are given in detail in Ref.
\cite{BabuWilczekPati}. Here, I will present only the results.  It
is found that  the four amplitudes $\widehat{A}_{\rm
std}(\overline{\nu}_\tau K^{+})$, $\widehat{A}_{\rm
std}(\overline{\nu}_{\mu}K^{+})$, $\widehat{A}_{\rm
new}(\overline{\nu}_{\tau}K^{+})$ and $\widehat{A}_{\rm
new}(\overline{\nu}_{\mu}K^{+})$ are in fact very comparable to
each other, within about a factor of two to five, either way.
Since there is no reason to expect a near cancellation between the
standard and the new operators, especially for both
$\overline{\nu}_{\tau}K^{+}$ and $\overline{\nu}_{\mu}K^{+}$
modes, we expect the net amplitude (standard + new) to be in the
range exhibited by either one. Following Ref.
\cite{BabuWilczekPati}, I therefore present the contributions from
the standard and the new operators separately.

One important consequence of the doublet-triplet splitting
mechanism for SO(10) outlined briefly in the appendix and in more
detail in Ref. \cite{BabuWilczekPati} is that the standard $d=5$
proton decay operators become inversely proportional to $M_{\rm
eff}\equiv [\lambda\langle\mathbf{45}_H\rangle]^2/$ $ M_{10'} \sim
M_{X}^2/M_{10'}$, rather than to $M_{H_C}$.  Here, $M_{10'}$
represents the mass of $\mathbf{10}_H'$, that enters into the D-T
splitting mechanism through effective coupling $\lambda
\mathbf{10}_H \mathbf{45}_H \mathbf{10}_H'$ in the superpotential
[see Appendix, Eq. (\ref{a1})].  As noted in Ref.
\cite{BabuWilczekPati}, $M_{10'}$ can be naturally suppressed (due
to flavor symmetries) compared to $M_X$, and thus $M_{\rm eff}$
correspondingly larger than $M_X$ by even one to three orders of
magnitude.  It should be stressed that $M_{\rm eff}$ does not
represent the physical masses of the color triplets or of the
other particles in the theory.  It is simply a parameter of order
$M_X^2/M_{10'}$.  {\it Thus values of $M_{\rm eff}$, close to or
even exceeding the Planck scale, do not in any way imply large
corrections from quantum gravity}.  Now accompanying the
suppression due to $M_{\rm eff}$, the standard proton decay
amplitudes for SO(10) possess an intrinsic enhancement as well,
compared to those for SU(5), owing primarily due to differences in
their Yukawa couplings for the up sector (see Appendix C of Ref.
\cite{BabuWilczekPati}).  As a result of this enhancement,
combined with the suppression due to higher values of $M_{\rm
eff}$, a typical standard $d=5$ amplitude for SO(10) is given by
(see Appendix C of Ref. \cite{BabuWilczekPati})
\begin{displaymath} \widehat{A}(\bar{\nu}_\mu
K^+)_{std}^{SO(10)}\approx(h_{33}^2/M_{\rm eff}) (2\times10^{-5}),
\end{displaymath} which should be compared with $\widehat{A}(\bar{\nu}_\mu
K^+)_{std}^{SU(5)}\approx(1.9\times10^{-8}) (\tan\beta/M_{H_C})$
[see Eq. (37)].  Note, taking $h_{33}^2\approx 1/4$, the ratio of
a typical SO(10) over SU(5) amplitude is given by \break
$(M_{H_c}/M_{\rm eff})(88)(3/\tan\beta)$.  Thus the enhancement by
a factor of about 88 (for $\tan\beta=3$), of the SO(10) compared
to the SU(5) amplitude, is compensated in part by the suppression
that arises from $M_{\rm eff}$ being larger than $M_{H_c}$.

In addition, note that in contrast to the case of SU(5), the
SO(10) amplitude does not depend {\it explicitly} on $\tan\beta$.
The reason is this: if the fermions acquire masses only through
the $\mathbf{10}_H$ in SO(10), as is well known, the up and down
quark Yukawa couplings will be equal. By itself, it would lead to
a large value of $\tan\beta=m_t/m_b\approx 60$ and thereby to a
large enhancement in proton decay amplitude. Furthermore, it would
also lead to the bad relations: $m_c/m_s=m_t/m_b$ and $V_{CKM}=1$.
However, in the presence of additional Higgs multiplets, in
particular with the mixing of $(\mathbf{16}_H)_d$ with
$\mathbf{10}_H$ (see Appendix and Section \ref{Understanding}),
(a) $\tan\beta$ can get lowered to values like 3-20, (b) fermion
masses get contributions from both $\langle\mathbf{16}_H
\rangle_d$ and $\langle\mathbf{10}_H \rangle$, which correct all
the bad relations stated above, and simultaneously (c) the
explicit dependence of $\widehat{A}$ on $\tan\beta$ disappears. It
reappears, however, through restriction on threshold corrections,
discussed below.

Although $M_{\rm eff}$ can far exceed $M_X$, it still gets bounded
from above by demanding that coupling unification, as observed
\footnote{ For instance, in the absence of GUT-scale threshold
corrections, the MSSM value of $\alpha_3(m_Z)_{MSSM}$, assuming
coupling unification, is given by
$\alpha_3(m_Z)_{MSSM}^\circ=0.125\pm 0.13$ \cite{Langacker}, which
is about 5-8\% higher than the observed value:
$\alpha_3(m_Z)_{MSSM}^\circ=0.118 \pm 0.003$
\cite{ParticleDataGroup}. We demand that this discrepancy should
be accounted for accurately by a net {\it negative} contribution
from D-T splitting and from ``other'' threshold corrections [see
Appendix, Eq. (\ref{a4})], without involving large cancellations.
That in fact does happen for the minimal Higgs system $(45, 16,
\overline{16})$ (see Ref. \cite{BabuWilczekPati}).  }, should
emerge as a natural prediction of the theory as opposed to being
fortuitous.  That in turn requires that there be no large
(unpredicted) cancellation between GUT-scale threshold corrections
to the gauge couplings that arise from splittings within different
multiplets as well as from Planck scale physics.  Following this
point of view, we have argued (see Appendix) that the net
``other'' threshold corrections to $\alpha_3(m_Z)$ arising from
the Higgs (in our case $\mathbf{45}_H$, $\mathbf{16}_H$ and
$\overline\mathbf{16}_H$) and the gauge multiplets should be
negative, but conservatively and quite plausibly no more than
about 10\%, at the electroweak scale. This in turn restricts how
big can be the threshold corrections to $\alpha_3(m_Z)$ that arise
from (D-T) splitting (which is positive).  Since the latter is
proportional to $\ln(M_{\rm eff}\cos\gamma/M_{X})$ (see Appendix),
we thus obtain an upper limit on $M_{\rm eff}\cos\gamma$. For the
simplest model of D-T splitting presented in Ref.
\cite{BabuWilczekPati} and in the Appendix [Eq. (\ref{a1})], one
obtains: $\cos\gamma\approx(\tan\beta)/(m_t/m_b)$.  An upper limit
on $M_{\rm eff}\cos\gamma$ thus provides an upper limit on $M_{\rm
eff}$ which is inversely proportional to $\tan{\beta}$. In short,
our demand of natural coupling unification, together with the
simplest model of D-T splitting, introduces an implicit dependence
on $\tan\beta$ into the lower limit of the SO(10)-amplitude---i.e.
$\widehat{A}(SO(10))\propto 1/M_{\rm eff} \geq$ [(a quantity)
$\propto \tan\beta$].  These considerations are reflected in the
results given below.

 Assuming $\tan{\beta}\ge 3$ and accurate coupling unification
(as described above), one obtains for the case of MSSM, a
conservative upper limit on $M_{\rm eff} \le 2.7\times 10^{18}$
GeV $(3/\tan\beta)$ (see Appendix and Ref.
\cite{BabuWilczekPati}). Using this upper limit, we obtain a lower
limit for the standard proton decay amplitude given by
\begin{equation}
\widehat{A}(\overline{\nu}_{\tau}K^{+})_{std}\geq
\left[\begin{array}{c}
  {(7.8\times10^{-24}\,\mbox{GeV}^{-1})\,(1/6\,\mbox{-\,1/4)}
    \,\,\,\,\,\,\,\,\mbox{case\ I}}\\
{(3.3\times10^{-24}\,\mbox{GeV}^{-1})\,(1/6\,\mbox{-\,1/2)}
\,\,\,\,\,\,\,\,\mbox{case\ II}}\end{array}\right]
\left(\begin{array}{c}\mbox{SO(10)/MSSM, with}\\
     \tan\beta \geq 3 \end{array}\right).
\label{e45}
\end{equation}
Substituting into Eq. (\ref{e40}) and adding the contribution from
the second  competing mode $\overline{\nu}_{\mu}K^{+}$, with a
typical branching ratio $R\approx0.3$, we obtain
\begin{equation}
\Gamma^{-1}(\overline{\nu}K^{+})_{std} \leq \left[\begin{array}{c}
{(0.18\times10^{31}\,\mbox{years})\,(1.6\,\mbox{-}\,0.7)}\\
{(0.4\times10^{31}\,\mbox{years})\,(4\,\mbox{-}\,0.44)}\end{array}\right]
\,\{64\,\mbox{-}\,1/64\}
 \left(\begin{array}{c}\mbox{SO(10)/MSSM, with}\\
     \tan\beta \geq 3 \end{array}\right).
\label{e46}
\end{equation}
The upper and lower entries in Eqs. (\ref{e45}) and (\ref{e46})
correspond to the cases I and II of the fermion mass-matrix with
the {\it extreme values} of $\epsilon'$---i.e.
$\epsilon'=2\times10^{-4}$ and $\epsilon'=0$---respectively, (see
Eq. (\ref{e33})).  The  uncertainty shown inside the square
brackets correspond to that in the  relative phases of the
different contributions.  The uncertainty of \{64 to 1/64\} arises
from that in $\beta_{H}$,  $(m_{\tilde{W}}/m_{\tilde{q}})$ and
$m_{\tilde{q}}$ [see Eq. (\ref{e39})]. Thus we find that for MSSM
embedded in SO(10), for the two extreme values of $\epsilon'$
(cases I and II) as mentioned above, the inverse partial proton
decay rate  should satisfy:
\begin{eqnarray}
\Gamma^{-1}(p\,\rightarrow\overline{\nu}K^{+})_{std} &\leq&
\left[\begin{array}{c}{0.20\times10^{31_{-1.7}^{+2.0}}\,\mbox{years}}\\
{0.32\times10^{31^{+2.4}_{-1.86}}\,\mbox
years}\end{array}\right]\nonumber \\
&\leq& \left[\begin{array}{c}{0.2\times10^{33}\,\mbox{years}}\\
{1\times10^{33}\,\mbox{years}}\end{array}\right]\,\,\,\,
 \left(\begin{array}{c}\mbox{SO(10)/MSSM, with}\\
    \tan\beta \geq 3 \end{array}\right)\,.
\label{e47}
\end{eqnarray}
The central value of the upper limit in Eq. (\ref{e47})
corresponds to taking the upper limit on $M_{\rm eff}\leq
2.7\times10^{18}$ GeV, which is obtained by restricting threshold
corrections as described above (and in the Appendix) and by
setting (conservatively) $\tan\beta\geq 3$. The uncertainties of
matrix element, spectrum and choice of phases are reflected in the
exponents. The uncertainty in the most sensitive entry of the
fermion mass matrix---i.e. $\epsilon'$---is incorporated (as
regards obtaining an upper limit on the lifetime) by going from
case I (with $\epsilon'=2\times 10^{-4}$) to case II
($\epsilon'=0$). Note that this increases the lifetime by almost a
factor of six. Any non-vanishing intermediate value of $\epsilon'$
would only shorten the lifetime compared to case II. In this
sense, the larger of the two upper limits quoted above is rather
conservative. We see that the predicted upper limit for case I of
MSSM (with the extreme value of $\epsilon'=2\times 10^{-4}$) is
lower than the empirical lower limit [Eq. (\ref{e42})] by a factor
of ten, while that for case II, i.e. $\epsilon'=0$ (with all the
uncertainties stretched as mentioned above) is about two times
lower than the empirical lower limit.

Thus the case of MSSM embedded in SO(10) is already tightly
constrained, to the point of being disfavored, by the limit on
proton lifetime.  The constraint is of course augmented especially
by {\it our requirement of natural coupling unification} which
prohibits accidental large cancellation between different
threshold corrections\footnote{Other authors (see e.g., Ref.
\cite{LucasRaby}) have considered proton decay in SUSY SO(10) by
allowing for rather large GUT-scale threshold corrections, which
do not, however, go well with our requirement of ``natural
coupling unification".} (see Appendix); and it will be even more
severe, especially within the simplest mechanism of D-T splitting
(as discussed in the Appendix), if $\tan\beta$ turns out to be
larger than 5 (say).  On the positive side, improvement in the
current limit by a factor of even 2 to 3 ought to reveal proton
decay, otherwise the case of MSSM embedded in SO(10), would be
clearly excluded.

\subsection{The case of ESSM}
\label{sec:essm}

Before discussing the contribution of the new $d=5$ operators to
proton decay, an interesting possibility, mentioned in the
introduction (and in footnote 16), that would be especially
relevant in the context of proton decay, if $\tan\beta$ is large,
is worth noting.  This is the case of the extended supersymmetric
standard model (ESSM), which introduces an extra pair of
vector-like families [$\mathbf{16}+\overline{\mathbf{16}}$ of
SO(10)], at the TeV scale \cite{BabuJi,BabuPatiStrem}.  Adding
such complete SO(10)-multiplets would of course preserve coupling
unification.  From the point of view of adding extra families,
ESSM seems to be the minimal and also the maximal extension of the
MSSM, that is allowed in that it is compatible with (a) LEP
neutrino-counting, (b) precision electroweak tests, as well as (c)
a semi-perturbative as opposed to non-perturbative gauge coupling
unification \cite{BabuJi,BabuPatiStrem}.  \footnote{ For instance,
addition of {\it two} pairs of vector-like families at the
TeV-scale, to the three chiral families, would cause gauge
couplings to become non-perturbative below the unification scale.
} {\it The existence of two extra vector-like families of quarks
and leptons can of course be tested at the LHC.}

Theoretical motivations for the case of ESSM arise on several
grounds: (a) it provides a better chance for stabilizing the
dilaton by having a semi-perturbative value for
$\alpha_{\mbox{\scriptsize{unif}}}\approx$ 0.35-0.3 \cite{BabuJi},
in contrast to a very weak value of 0.04 for MSSM; (b) owing to
increased two-loop effects \cite{BabuJi,KoldaRussell}, it raises
the unification scale $M_X$ to $(1/2-2)\times 10^{17}$ GeV and
thereby considerably reduces the problem of a mismatch
\cite{DienesJCP} between the MSSM and the string unification
scales (see Section  \ref{Need}); (c) It lowers the GUT-prediction
for $\alpha_3(m_Z)$ to (0.112--0.118) (in absence of
unification-scale threshold corrections), which is in better
agreement with the data than the corresponding value of
(0.125--0.13) for MSSM; and (d) it provides a simple reason for
inter-family mass-hierarchy \cite{BabuJi,BabuPatiStrem}.  In this
sense, ESSM, though less economical than MSSM, offers some
distinct advantages.

In the present context, because of (b) and (c), ESSM naturally
enhances the GUT-prediction for proton lifetime, in full accord
with the data \cite{SKlimit}.  As explained in the appendix, the
net result of these two effects---i.e. a raising of $M_X$ and a
lowering of $\alpha_3(m_Z)_{\mbox{\scriptsize{ESSM}}}^\circ$---is
that for ESSM embedded in SO(10), $\tan\beta$ can span a wide
range from 3 to even 30, and simultaneously the value or the upper
limit on $M_{\rm eff}$ can range from $(60$ to $6)\times 10^{18}$
GeV, in full accord with our criterion for accurate coupling
unification discussed above.

As a result, in contrast to MSSM, ESSM allows for larger values of
$\tan\beta$ (like 10 or 20), without needing large threshold
corrections, and simultaneously without conflicting with the limit on
proton lifetime.

To be specific, consider first the case of a moderately large
$\tan\beta=10$ (say), for which one obtains $M_{\rm eff}\approx
1.8\times 10^{19}$ GeV, with the ``other'' threshold correction
$-\delta_3'$ being about 5\% (see Appendix for definition).  In
this case, one obtains:
\begin{equation}
\Gamma^{-1}(\overline{\nu}K^{+})_{\rm std} \approx
\left[\begin{array}{c}{(1.6-0.7)}\\
{(10-1)}\end{array}\right]\,\{64-1/64\}\, (7\times
10^{31}\,\mbox{years})
\left(\begin{array}{c}\mbox{SO(10)/ESSM, with} \\
\tan\beta=10\end{array}\right).  \label{e49-2}
\end{equation}
As before, the upper and lower entries correspond to cases I
($\epsilon'=2\times 10^{-4}$) and II ($\epsilon'=0$) of the
fermion mass-matrix [see Eq. (\ref{e33})].  The uncertainty in the
upper and lower entries in the square bracket of Eq. (\ref{e49-2})
corresponds to that in the relative phases of the different
contributions for the cases I and II respectively, while the
factor \{64-1/64\} corresponds to uncertainties in the SUSY
spectrum and the matrix element (see Eq.  ({\ref{e39}})).

We see that by allowing for an uncertainty of a factor of
$(30-100)$ jointly from the two brackets proton lifetime arising
from the standard operators would be expected to lie in the range
of $(2.1-7)\times10^{33}$ years, for the case of ESSM embedded in
SO(10), even for a moderately large $\tan\beta=10$.  Such a range
is compatible with present limits, but accessible to searches in
the near future.

The other most important feature of ESSM is that, by allowing for
larger values of $M_{\rm eff}$, especially for smaller values of
$\tan\beta\approx 3$ to $5$ (say), {\it the contribution of the
standard operators by itself can be perfectly consistent with
present limit on proton lifetime even for almost central or
``median'' values of the parameters pertaining to the SUSY
spectrum, the relevant matrix element, $\epsilon'$ and the
phase-dependent factor}.

For instance, for ESSM, one obtains $M_{\rm eff}\approx (4.5\times
10^{19} \mbox{GeV}) (4/\tan\beta)$, with the ``other" threshold
correction - $\delta_3'$ being about 5\% [see Appendix and Eq.
(\ref{MeffEq})]. Now, {\it combining} cases I ($\epsilon'=2\times
10^{-4}$) and II ($\epsilon'=0$), we see that the square bracket
in Eq. (\ref{e49-2}) which we will denote by [S], varies from 0.7
to 10, depending upon the relative phases of the different
contributions and the values of $\epsilon'$. Thus as a ``median''
value, we will take $[S]_{\rm med} \approx$ 2 to 6. The curly
bracket \{64-1/64\}, to be denoted by \{C\}, represents the
uncertainty in the SUSY spectrum and the matrix element [see Eq.
(\ref{e39})]. Again as a ``nearly central" or ``median" value, we
will take $\{C\}_{\rm med}\approx 1/6 \mbox{ to } 6$. Setting
$M_{\rm eff}$ as above we obtain
\begin{equation}\label{e50-2}
  \Gamma^{-1}(\bar{\nu}K^+)_{\rm std}^{\rm ``median"} \approx
    [S]_{\rm med}\{C\}_{\rm med} (0.45\times 10^{33} \mbox{years})
    (4/\tan\beta)^2 (\mbox{SO(10)/ESSM}).
\end{equation}
Choosing a few sample values of the effective parameters [S] and
\{C\}, with low values of $\tan\beta = 3$ to 5, the corresponding
values of $\Gamma^{-1}(\bar{\nu}K^+)$, following from Eq.
(\ref{e50-2}), are listed below in Table~\ref{tab:1}.

Note that ignoring contributions from the new $d=5$ operators for
a moment\footnote{As I will discuss in the next section, we of
course expect the new $d=5$ operators to be important and
significantly influence proton lifetime (see e.g.
Table~\ref{tab:2}). Entries in Table~\ref{tab:1} could still
represent the actual expected values of proton lifetimes, however,
if the parameter K defined in \ref{Preliminaries} (also see
\ref{Contribution}) happens to be unexpectedly small ($\ll 1$). },
the entries in Table~\ref{tab:1} represent {\it a very plausible
range of values} for the proton lifetime, for the case of ESSM
embedded in SO(10), with $\tan\beta\approx 3 \mbox{ to } 5$ (say),
{\it rather than upper limits for the same}. This is because they
are obtained for ``nearly central" or ``median" values of the
parameters represented by the values of [S] and \{C\}, as
discussed above. For instance, consider the cases \{C\}=1 and
\{C\}=1/2 respectively, which (as may be inferred from the table)
can quite plausibly yield proton lifetimes in the range of (2 to
5)$\times 10^{33}$ years Now \{C\}=1 corresponds, e.g., to
$\beta_H=0.014$ GeV$^3$ (the central value of Ref. \cite{Aoki})
$m_{\tilde{q}}=1.2$ TeV and $m_{\tilde{W}}/m_{\tilde{q}}=1/6$ [see
Eq. (\ref{e39})], while that of \{C\}=1/2 would correspond, for
example, to $\beta_H=0.014$ GeV$^3$, with $m_{\tilde{q}}\approx
710$ GeV and $m_{\tilde{W}}/m_{\tilde{q}}\approx 1/6$. {\it In
short, for the case of ESSM, with low values of $\tan\beta\approx$
3 to 5 (say), squark masses can be well below 1 TeV, without
conflicting with present limit on proton lifetime}. This feature
is not permissible within MSSM embedded in SO(10).

\begin{table}[htb]
{\bf \caption[*]{\bf \large Proton lifetime, based on
contributions from only the standard operators for the case of
ESSM embedded in SO(10), with
 parameters being in the ``median'' range.\label{tab:1}}}\medskip
\begin{tabular}{|c|c|c|c|}
\hline
  $\tan\beta=3$ & $\tan\beta=3$ & $\tan\beta=5$ & $\tan\beta=5$ \\
  \hspace{0pt}[S]=3 & [S]=6 & [S]=5.4   & [S]=6 \\
  \{C\}=1/2 to 4& \{C\}=1/2 to 1& \{C\}=1 to 6  & \{C\}=1 to 4  \\
\hline
  $\Gamma^{-1}(\bar{\nu}K^+)_{ESSM}^{std}\approx$ &
    $\Gamma^{-1}(\bar{\nu}K^+)_{ESSM}^{std}\approx$ &
        $\Gamma^{-1}(\bar{\nu}K^+)_{ESSM}^{std}\approx$ &
        $\Gamma^{-1}(\bar{\nu}K^+)_{ESSM}^{std}\approx$   \\
 $(1.2\mbox{ to }10)\times 10^{33}$ yrs &
    $(2.5\mbox{ to }5)\times 10^{33}$ yrs &
    $(1.6\mbox{ to }10)\times 10^{33}$ yrs &
    $(1.8\mbox{ to }7.3)\times 10^{33}$ yrs \\
\hline
\end{tabular}
\end{table}

Thus, confining for a moment to the standard operators only, if
ESSM represents low-energy physics, and if $\tan\beta$ is rather
small (3 to 5, say), we do not have to stretch the uncertainties
in the SUSY spectrum and the matrix elements to their extreme
values (in contrast to the case of MSSM) in order to understand
why proton decay has not been seen as yet, and still can be
optimistic that it ought to be discovered in the near future, with
a lifetime $\leq 10^{34}$ years.  The results for a wider
variation of the parameters are listed in Table~\ref{tab:2}, where
contributions of the new $d=5$ operators are also shown.

It should also be remarked that if in the unlikely event, all the
parameters (i.e.  $\beta_H$, $(m_{\tilde{W}}/m_{\tilde{q}})$,
$m_{\tilde{q}}$ and the phase-dependent factor) happen to be
closer to their extreme values so as to extend proton lifetime,
and if $\tan\beta$ is small ($\approx 3$ to 5, say) and at the
same time the value of $M_{\rm eff}$ is close to its allowed upper
limit (see Appendix), the standard $d=5$ operators by themselves
would tend to yield proton lifetimes exceeding even (0.8 to
2.5)$\times10^{34}$ years for the case of ESSM, (see Eq.
(\ref{e49-2}) and Table~\ref{tab:2}).  In this case (with the
parameters having nearly extreme values), however, as I will
discuss shortly, the contribution of the new $d=5$ operators
related to neutrino masses [see Eq. (\ref{e35})], are likely to
dominate and quite naturally yield lifetimes bounded above in the
range of $(1-10)\times 10^{33}$ years (see Section
\ref{Contribution} and Table~\ref{tab:2}).  {\it Thus in the
presence of the new operators, the range of $(10^{33}-10^{34})$
years for proton lifetime is not only very plausible but it also
provides a reasonable upper limit, for the case of ESSM embedded
in SO(10)}.

\subsection{Contribution from the new d=5 operators}\label{Contribution}

As mentioned in Section \ref{Preliminaries}, for supersymmetric
G(224)/SO(10), there very likely exists a new set of $d=5$
operators, related to neutrino masses, which can induce proton
decay [see Eq. (\ref{e41})].  The decay amplitude for these
operators for the leading mode (which in this case is
$\bar{\nu}_\mu K^+$) becomes proportional to the quantity $P\equiv
\{(f_{33})_\nu\langle \overline{\mathbf{16}}_H\rangle/M\}
h_{33}K/(M_{16}\tan\gamma)$, where $(f_{33})_\nu$ and $h_{33}$ are
the effective couplings defined in Eqs. (\ref{e7}) and (\ref{e15})
respectively, and $M_{16}$ and $\tan\gamma$ are defined in the
Appendix.  The factor K, defined by $(f_{33})_p\equiv (f_{33})_\nu
K$, is expected to be of order unity (see Section
\ref{Preliminaries} for the origin of K).  As a plausible range,
we would take $K\approx 1/5$ to 2. Using
$M_{16}\tan\gamma=\lambda'\langle\overline{\mathbf{16}}_H\rangle$
(see Appendix), and $h_{33}\approx 1/2$ (given by top mass), one
gets: $P\approx [(f_{33})_\nu/M](1/2\lambda')K$.  Here M denotes
the string or the Planck scale (see Section  \ref{Mass} and
footnote 2); thus $M\approx (1/2 -1)\times 10^{18}$ GeV; and
$\lambda'$ is a quartic coupling defined in the appendix. Validity
of perturbative calculation suggests that $\lambda'$ should not
much exceed unity, while other considerations suggest that
$\lambda'$ should not be much less than unity either (see Ref.
\cite{BabuWilczekPati}, Section 6 E).  Thus, a plausible range for
$\lambda'$ is given by $\lambda'\approx(1/2 -\sqrt{2})$. (Note it
is only the upper limit on $\lambda'$ that is relevant to
obtaining an upper limit on proton lifetime).  Finally, from
consideration of $\nu_\tau$ mass, we have $(f_{33})_\nu\approx 1$
(see Section  \ref{Mass}). We thus obtain:  $P\approx
(5\times10^{-19}\mbox{GeV}^{-1})(1/\sqrt{2}$ to 4)K. Incorporating
a further uncertainty by a factor of (1/2 to 2) that arises due to
choice of the relative phases of the different contributions (see
Ref. \cite{BabuWilczekPati}), the effective amplitude for the new
operator is given by
\begin{equation}\label{e51} \widehat{A}(\bar{\nu}_\mu K^+)_{\rm
new}\approx (1.5\times 10^{-24} \mbox{GeV}^{-1}) (1/2\sqrt{2}\,\,
\mbox{to}\,\, 8)K \end{equation} Note that this new contribution
is independent of $M_{\rm eff}$; {\it thus it is the same for ESSM
as it is for MSSM, and it is independent of $\tan\beta$}.
Furthermore, it turns out that the new contribution is also
insensitive to $\epsilon'$; thus it is nearly the same for cases I
and II of the fermion mass-matrix.  Comparing Eq.  (\ref{e51})
with Eq. (\ref{e45}) we see that the new and the standard
operators are typically quite comparable to one another.  Since
there is no reason to expect near cancellation between them
(especially for both $\bar{\nu}_\mu K^+$ and $\bar{\nu}_\tau K^+$
modes), we expect the net amplitude (standard+new) to be in the
range exhibited by either one.  It is thus useful to obtain the
inverse decay rate assuming as if the new operator dominates.
Substituting Eq.  (\ref{e51}) into Eq.  (\ref{e40}) and allowing
for the presence of the $\bar{\nu}_\tau K^+$ mode with an
estimated branching ratio of nearly 0.4 (see Ref.
\cite{BabuWilczekPati}), one obtains
\begin{equation}
\Gamma^{-1}(\overline{\nu}K^{+})_{\rm
new}\,\approx\,(0.25\times10^{31}\,
\mbox{years})\,[8\,\mbox{-}\,1/64]\,\{64\,\mbox{-}\,1/64\}
(K^{-2}\approx 25\,\,\mbox{to}\,\, 1/4)\,.  \label{e52}
\end{equation}
The square bracket represents the uncertainty reflected in Eq.
(\ref{e51}), while the curly bracket corresponds to that in the
SUSY spectrum and matrix element (Eq.  (\ref{e39})).  Allowing for
the net uncertainty factor at the upper end, arising jointly from
the {\it three brackets} in Eq. (\ref{e52}) to be 1000 to 4000
(say), which can be realized for plausible range of values of the
parameters (see below), the new operators related to neutrino
masses, by themselves, lead to a proton decay lifetime given by:
\begin{equation}
\Gamma^{-1}(\overline{\nu}K^{+})^{\rm upper}_{\rm new}\, \approx
(2.5\,\mbox{-}\,10)\times10^{33}\,\mbox{years}\  \mbox{(SO(10) or
string G(224))}\ (\mbox{Indep. of }\tan\beta)\ . \label{e53}
\end{equation}
The superscript ``upper'' corresponds to estimated lifetimes near
the upper end. For instance, taking the curly bracket in Eq.
(\ref{e52}) to be $\approx 8$ to 16 (say) [corresponding for
example, to $\beta_H=0.010$ GeV$^3$,
$(m_{\tilde{W}}/m_{\tilde{q}})\approx 1/12$ and
$m_{\tilde{q}}\approx (1$ to 1.4)(1.2 TeV)], instead of its
extreme value of 64, and setting the square bracket in Eq.
(\ref{e52}) to be $\approx 6$, and $K^{-2}\approx 20$, which are
quite plausible, we obtain: $\Gamma^{-1}(\bar{\nu}K^+)_{\rm
new}\approx (2.5-5)\times 10^{33}$ years; independently of
$\tan\beta$, for both MSSM and ESSM. Proton lifetime for other
choices of parameters, which lead to similar conclusion, are
listed in Table~\ref{tab:2}.

It should be stressed that the standard $d=5$ operators [mediated
by the color-triplets in the $10_H$ of SO(10)] may naturally be
absent for a string-derived G(224)-model (see e.g.  Ref.
\cite{Antoniadis} and \cite{FaraggiHalyo}), but the new $d=5$
operators, related to the Majorana masses of the RH neutrinos and
the CKM mixings, should very likely be present for such a model,
as much as for SO(10).  These would induce proton decay \footnote{
In addition, quantum gravity induced $d=5$ operators are also
expected to be present at some level, depending upon the degree of
suppression of these operators due to flavor symmetries (see e.g.
Ref. \cite{JCPProton}).  }. {\it Thus our expectations for the
proton decay lifetime [as shown in Eq.  (\ref{e53})] and the
prominence of the $\mu^{+}K^{0}$ mode (see below) hold for a
string-derived G(224)-model, just as they do for SO(10)}.  For a
string - G(224) - model, however, {\it the new d=5 operators would
be essentially the sole source of proton decay}$^{21}$.

Nearly the same situation emerges for the case of ESSM embedded in
G(224) or SO(10), with low $\tan\beta (\approx 3$ to 10, say),
especially if the parameters (including $\beta_H$,
$m_{\tilde{W}}/m_{\tilde{q}}$, $m_{\tilde{q}}$, the
phase-dependent factor as well as $M_{\rm eff}$) happen to be
somewhat closer to their extreme values so as to extend proton
lifetime.  In this case, (that is for ESSM) as noted in the
previous sub-section, the contribution of the standard $d=5$
operators would be suppressed; and proton decay would proceed
primarily via the new operators with a lifetime quite plausibly in
the range of $10^{33}-10^{34}$ years, as exhibited above.

\subsection{The Charged Lepton Decay Modes \boldmath
$(p\rightarrow\mu^{+}K^{0}$ and $p\rightarrow e^+\pi^0$)
\unboldmath} \label{TheCharged}

I now note a distinguishing feature of the SO(10) or the
G(224) model presented here.
Allowing for  uncertainties in the way the standard and the
new operators can combine  with each other for the three leading modes
i.e. $\overline{\nu}_{\tau}K^{+}$,  $\overline{\nu}_{\mu}K^{+}$ and
$\mu^{+}K^{0}$, we obtain (see Ref. \cite{BabuWilczekPati} for  details):
\begin{equation}
B(\mu^{+}K^{0})_{std+new}\,\approx\,\left[1\%\,\,
\mbox{to}\,\,50\%\right]\,\kappa\,\,\,\,\mbox{(SO(10)
or string G(224))}
\label{e54}
\end{equation}
where $\kappa$ denotes the ratio of the squares of relevant matrix
elements  for the $\mu^{+}K^{0}$ and $\overline{\nu}K^{+}$ modes. In
the absence of a reliable lattice calculation for the $\bar{\nu}K^{+}$
mode, one should remain open to the possibility  of
$\kappa\approx1/2$ to 1 (say).
 We find that for a
large range of parameters, the branching ratio $B(\mu^{+}K^{0})$ can
lie in the range of 20 to 40\% (if $\kappa \approx1$). This prominence of
the $\mu^{+}K^{0}$ mode for the SO(10)/G(224) model
 is primarily due to contributions from the new $d=5$
operators. This contrasts sharply with the minimal SU(5) model, in
which the $\mu^{+}K^{0}$ mode is expected to have a branching ratio of
only about $10^{-3}$. In short, prominence of the $\mu^{+}K^{0}$ mode,
if seen, would clearly show the relevance of the new operators, and
thereby reveal the proposed link between neutrino masses and proton
decay \cite{BPW1}.

The $d=5$ operators as described here (standard and new) would
lead to highly suppressed $e^+\pi^0$ mode, for MSSM or ESSM
embedded in SO(10). The gauge boson-mediated $d=6$ operators,
however, still give (using the recently determined matrix element
$\alpha_H=0.015\pm 0.001$ GeV$^3$ \cite{Aoki}) proton decaying
into $e^+\pi^0$ with an inverse rate:
\begin{equation}
\label{eq55} \Gamma^{-1}(p\rightarrow e^+\pi^0)_{\rm MSSM}^{\rm
SO(10)/SU(5)}\approx 10^{35\pm 1} \mbox{years} \ .
\end{equation}
This can well be as short as about $10^{34}$ years.  For the case
of ESSM embedded into SO(10) [or for an analogous case embedded
into SU(5)], there are two new features. Considering that in this
case, both $\alpha_{\rm unif}$ and the unification scale $M_X$
(thereby the mass $M_V$ of the $(X,Y)$ gauge bosons) are raised by
nearly a factor of (6 to 7) and (2.5 to 5) respectively, compared
to those for MSSM (see discussions in Section \ref{sec:essm}), and
that the inverse decay rate is proportional to
$(M_V^4/\alpha^2_{\rm unif})$, we expect
\begin{equation}
\label{eq56} \Gamma^{-1}(p\rightarrow e^+\pi^0)_{\rm ESSM}^{\rm
SO(10)/SU(5)} \approx (1\mbox{ to } 17)\Gamma^{-1}(p\rightarrow
e^+\pi^0)_{\rm MSSM} ^{\rm SO(10)/SU(5)} \ .
\end{equation}
The net upshot is that the gauge boson-mediated $d=6$ operators
can quite plausibly lead to observable $e^+\pi^0$ decay mode with
an inverse decay rate in the range of 10$^{34}$-10$^{35}$ years.
For ESSM embedded in SO(10), there can be the interesting
situation that both $\bar{\nu}K^+$ (arising from $d=5$) and
$e^+\pi^0$ (arising from $d=6$) may have comparable rates, with
proton having a lifetime $\sim (1/2$-2) $\times 10^{34}$ years. It
should be stressed that the $e^+\pi^0$-mode is the {\it common
denominator} of all GUT models (SU(5), SO(10), etc.) which unify
quarks and leptons and the three gauge forces. Its rate as
mentioned above is determined essentially by the SUSY
unification-scale, without the uncertainty of the SUSY-spectrum. I
should also mention that the $e^+\pi^0$-mode is predicted to be
the dominant mode in the flipped SU(5) $\times$ U(1)-model
\cite{ref:zz}. For these reasons, intensifying the search for the
$e^+\pi^0$-mode to the level of sensitivity of about $10^{35}$
years in the next generation proton decay detector should be well
worth the effort.

Before summarizing the results of this section, I note below a few
distinctive features of the conventional approach adopted here
compared to those of some alternatives.

\subsection{Conventional Versus Other Approaches}

In these lectures, as elaborated in Section 3, I have pursued
systematically the consequences for fermion masses, neutrino
oscillations {\it and} proton decay of the assumption that
essentially the conventional picture of SUSY grand unification
\cite{JCPAS,GG,GQW,SO(10),Langacker} holds, providing a good
effective theory in 4D between the conventional GUT-scale $M_X\sim
2\times 10^{16}$ GeV (for ESSM,
 $M_X\sim (1/2\mbox{-}2)\times 10^{17}$ GeV) and the conventional string
 scale $M_{\rm st}\sim (\mbox{few to 10})\times 10^{17}$ GeV. Believing in an
 underlying string/M-theory, and yet knowing that a preferred ground state of
 this theory is not yet in hand, the attitude, based on a bottom-up approach,
 has been to subject the assumed effective theory of grand unification to as
 many low-energy tests as possible, and to assess its soundness
 on empirical grounds. With this in mind, I have assumed that either a
 realistic 4D SO(10)-solution (with the desired mechanism of doublet-triplet
 splitting operating in 4D), or a suitable string-derived G(224)-solution
 (with $M_X\sim (1/2)M_{\rm st}$, see footnote 2) emerges effectively from an
 underlying string theory at the conventional string scale as mentioned above,
 and that the G(224)/SO(10) symmetry breaks into G(213) at the conventional
 GUT-scale $M_X$. The extra dimensions of string/M-theory are assumed to be
 tiny lying between the GUT-scale size $\sim M_X^{-1}$ and the string-size
 $M_{\rm st}^{-1}$, so as not to disturb the successes of GUT (see below).
 As mentioned before, this conventional picture of grand unification
 described above seems to be
 directly motivated on observational grounds such as those based on (a)
 coupling unification or equivalently the agreement between the observed
 and the predicted values of $\sin ^2\theta_W$ (see Section 3), (b)
 neutrino masses including $\Delta m^2(\nu_\mu\mbox{-}\nu_\tau)$ and
 (c) the fact that spontaneous violation of B--L local symmetry seems to be
 needed to implement baryogenesis via leptogenesis
 \cite{KuzminRubakov,LeptoB}. The relevance of the group theory of
 G(224)/SO(10)-symmetry for the 4D theory is further suggested by the success
 of the predictions of the masses and the mixings of all fermions
 including neutrinos; these include $m^0_b\approx m^0_\tau$,
 $m(\nu^\tau_{\rm Dirac})\approx m_t(M_X)$, and the smallness of
 $V_{cb}\approx 0.04$ correlated with the largeness of
 $\sin^22\theta^{\rm osc}_{\nu_\mu\nu_\tau}\approx 1$  (see Section 5).

 In contrast to this conventional approach based on a presumed string-unified
 G(224) or an SO(10)-symmetry, there are several alternative approaches
 (scenarios) which have been proposed in the literature in recent years. Of
 importance is the fact that in many of these alternatives an attempt is made
 to strongly suppress proton decay, in some cases exclusively the $d=5$ operators
 (though not necessarily the $d=6$), invariably utilizing a higher dimensional
 mechanism. Each of these alternatives is interesting in its own right.
 However, it seems to me that the collection of successes mentioned above is
 not (yet) realized within these alternatives. For comparison, I mention
 briefly only a few, leaving out many interesting variants.

One class of alternatives is based on the idea of  large extra
dimensions \cite{ref:36,largeextradim}, or a low string scale of
order one TeV \cite{ref:37}. Though most intriguing, it does not
seem to provide simple explanations for (a) coupling unification,
(b) neutrino-masses (or their (mass)$^2$-differences) of the
observed magnitudes\footnote{By placing the singlet (right-handed)
neutrino in the bulk, for example, one can get a light Dirac
neutrino \cite{ref:aba} with a mass $m_\nu\approx \kappa  v_{EW}
M^*/M_{\rm Pl}\approx \kappa(2\times10^{-5}$ eV), where
$M^*\approx 1$ TeV, $M_{\rm Pl}\approx 10^{19}$ GeV (as in
\cite{ref:aba}), and $\kappa$ is the effective Yukawa coupling. To
get $m_\nu\sim 1/20$ eV (for SuperK), one would, however, need too
large a $\kappa \sim 2\times 10^3$ and/or too large a value for
$M^*\gsim$ 100 TeV; which would seem to face the gauge-hierarchy
problem.}, (c) a large (or maximal) $\nu_\mu$-$\nu_\tau$
oscillation angle, and (d) baryogenesis via leptogenesis that
seems to require violation of B--L at high temperatures. Within
this scenario, quantum-gravity induced proton decay would
ordinarily be extra rapid.  This is prevented, for example, by
assuming that quarks and leptons live in different positions in
the extra dimension. It appears to me that this idea (introduced
just to prevent proton decay) however, sacrifices the simple
reason for the co-existence quarks and leptons that is provided by
a gauge unification of matter within a family as in G(224) or
SO(10).

There is an alternative class of attempts, carried out again in
the context of higher dimensional theories, which, in contrast to
the case mentioned above, assume that the extra dimensions (d
$>4$) are all small, lying between (or around) the conventional
GUT and string scales. The approach of this class of attempts is
rather close in spirit to that of  the conventional approach of
grand unification pursued here (see Section 3).  As may be seen
from the discussions below, they could essentially coincide with
the string-unified G(224)-picture presented here if the effective
symmetry in 4D, below the string (or compactification) scale,
contains at least the G(224) symmetry.

Motivated by the original attempts carried out in the context of
string theory \cite{Candelas} most of the recent attempts in the
class mentioned above are made in the spirit of a bottom-up
approach\footnote{This is of course also the case for the approach
adopted here which is outlined in Section 3.} to physics near the
GUT and the string scales. They assume, following the spirit of
the results of Ref. \cite{Candelas}, and of analogous results
obtained for the free fermionic formulation of string theory
\cite{Kawai} (for applications based on this formulation, see
e.g., Ref. \cite{Antoni}, \cite{Antoniadis}, \cite{FaraggiHalyo}
and \cite{Faraggi}), that grand unification occurs, through
symmetries like E$_6$, SO(10) or SU(5), only in some higher
dimension (d $>4$), and that the breaking of the unification gauge
symmetry to some lower symmetry containing the standard model
gauge group as well as doublet-triplet splitting occurs in the
process of compactification.  More specifically the latter two
phenomena take place through either (a) Wilson lines
\cite{Candelas}, or (b) orbifolds  \cite{Dixon} (for an incomplete
list of recent attempts based on orbifold compactification, see
e.g., Refs.
\cite{Kawamura,Hall,Altarelli,ref:79,Kakizari,Hall2,Asaka,Hall3,ref:85}),
or (c) essentially equivalently by a set of boundary conditions
together with the associated GSO projections  for the free
fermionic formulation (see e.g.,
\cite{Antoniadis,FaraggiHalyo,Antoni,Faraggi}), or (d) discrete
symmetries operating in higher dimensions \cite{Witten}.

Most of these attempts end up not only in achieving (a)
doublet-triplet splitting by projecting out the relevant color
triplets from the zero mode-spectrum in 4D, and (b) gauge symmetry
breaking, as mentioned above, but also (c) suppressing strongly or
eliminating the $d=5$ proton decay operators.  It should be
mentioned, however, that in some of these attempts (see e.g.,
\cite{Hall}), the mass of the $X$ gauge boson is suggested to be
lower than the conventional GUT-scale of 2$\times 10^{16}$ GeV by
about a factor of 3 to 8; correspondingly they raise the prospect
for observing the $d=6$ gauge boson mediated $e^+\pi^0$ mode,
which is allowed in \cite{Hall}.

One crucial distinction between the various cases is provided by
the nature of the effective gauge symmetry that is realized in 4D,
below the string (or compactification) scale. References
\cite{Kawamura,Hall,Altarelli,Kakizari,ref:xx,Hall2} assume a
supersymmetric SU(5) gauge symmetry in 5D, which is broken down to
the standard model gauge symmetry in 4D through compactification.
References \cite{Asaka} and \cite{Hall3}, on the other hand,
assume a supersymmetric SO(10) gauge symmetry in 6D and show
(interestingly enough) that there are two 5D subspaces containing
G(224) and SU(5)$\times$U(1) subgroups respectively, whose
intersection leads to
SU(3)$\times$SU(2)$\times$U(1)$_Y\times$U(1)$_X$ in 4D, which
contains B--L. [An alternative construction using SO(10) in 5D
leads just to the standard model gauge group in 4D \cite{ref:85}.]
While it is desirable to have B--L in 4D, consistent breaking of
U(1)$_X$ (or B--L) and generating desired masses of the right
handed neutrinos, not to mention the masses and the mixings of the
other fermions, is not yet realized in these constructions.

For comparison, it seems to me that at the very least B--L should
emerge as a generator in 4D to implement baryogenesis via
leptogenesis, and also to protect RH neutrinos from acquiring a
string-scale mass. This feature is not available in models which
start with SU(5) in 5D. Furthermore, the full SU(4)-color
symmetry, which of course contains B--L, plays a crucial role in
yielding not only $m^0_b\approx m^0_\tau$ but also (a)
$m(\nu^\tau_{\rm Dirac})\approx m_t(M_X)$ that is needed to
account for $m(\nu_\tau)$ or rather $\Delta
m^2(\nu_\mu\mbox{-}\nu_\tau)$, in accord with observation (see
Section 4), and (b) the smallness of $V_{cb}$ together with the
near maximality of $\sin^22\theta^{\rm osc}_{\nu_\mu\nu_\tau}$
(see Section  5).  The symmetry SU(2)$_L\times$SU(2)$_R$ is also
most useful in that it relates the masses and mixings of the up
and the down sectors.  Without such relations, we will not have
the predictivity of the framework presented in Section  5.

In short, as mentioned before, certain intriguing features of the
masses and mixings of all fermions including neutrinos, of the
type mentioned above, as well as the need for leptogenesis, seem
to strongly suggest that the effective symmetry below the
string-scale in 4D should contain minimally the symmetry G(224)
[or a close relative G(214)] and maximally SO(10).  The
G(224)/SO(10)-framework developed here has turned out to be the
most predictive, in large part by virtue of its group structure
and the assumption of minimality of the Higgs system.  Given that
it is also most successful so far, as regards its predictions,
derivation of such a picture from an underlying theory, especially
at least that based on an effective G(224)-symmetry\footnote{For
this case, following the examples of Refs.  \cite{Antoniadis} and
\cite{FaraggiHalyo}, the color triplets in the $\mathbf{10}_H$ of
SO(10) would be projected out of the zero-mode spectrum, and thus
the standard $d=5$ operators which would have been induced by the
exchange of such triplets would be absent, as in Refs.
\cite{Kawamura,Hall,Altarelli,Kakizari,ref:xx,Hall2,Asaka,Hall3,ref:85,Witten}.
But, as long as the Majorana masses of the RH neutrinos are
generated as in Section  4, the new neutrino-mass related $d=5$
proton decay operators would generically be present (see Section 6
E).} in 4D leading to the pattern of Yukawa couplings presented
here remains a challenge.\footnote{In this regard,
three-generation solutions containing the G(224)-symmetry in 4D
have been obtained in the context of the fermionic formulation of
string theory in Ref. \cite{Antoniadis}, within type-I string
vacua with or without supersymmetry in \cite{ref:a,ref:b,ref:c} in
the context of D-brane inspired models in \cite{ref:d}, within
type-I string-construction or string-motivated models obtained
from intersecting D-branes (with G(224) breaking into G(213) at
$M_X\sim M_{\rm st}$) in \cite{ref:e,ref:f}, in string model with
unification at the string scale in \cite{ref:g}, and in other
contexts (see e.g. \cite{ref:89} and \cite{ref:90}).} Pending such
a derivation, however, given the empirical support it has received
so far, it makes sense to test the supersymmetric
G(224)/SO(10)-framework, and thereby the {\em conventional picture
of grand unification} on which it rests, thoroughly. There are two
notable missing pieces of this picture. One is supersymmetry which
will be probed at the LHC and a future NLC. The other, that
constitutes the hallmark of grand unification, is proton decay.
The results of this section on proton decay are summarized below.

\subsection{Section Summary}

Given the importance of proton decay, a systematic study of this
process has been carried out within the supersymmetric
SO(10)/G(224)-framework\footnote{As described in Sections
\ref{Need}, \ref{Mass} and \ref{Understanding}.}, with special
attention paid to its dependence on fermion masses and threshold
effects.  A representative set of results corresponding to
different choices of parameters is presented in Tables~\ref{tab:1}
and \ref{tab:2}. Allowing for the ESSM-variant, the study strongly
suggests that an upper limit on proton lifetime is given by
\begin{equation}
\tau_{\rm
proton}\,\leq\,(1/3\,\mbox{-}\,2)\times10^{34}\,\,\mbox{years}\,,
\label{e51-2}
\end{equation}
with $\overline{\nu}K^{+}$ being the
dominant decay mode, and quite possibly $\mu^+K^{0}$ and
$e^+\pi^0$ being prominent.  Although
there are uncertainties in the matrix element, in the
SUSY-spectrum, in the phase-dependent factor, $\tan\beta$ and in
certain sensitive elements of the fermion mass matrix, notably
$\epsilon'$ (see Eq.  (\ref{e47}) for predictions in cases I
versus II), this upper limit is obtained, for the case of MSSM
embedded in SO(10), by allowing for a generous range in these
parameters and stretching all of them in the same direction so as
to extend proton lifetime.  In this sense, while the predicted
lifetime spans a wide range, the upper limit quoted above, in fact
more like 10$^{33}$ years, is most conservative, for the
case of MSSM (see Eq.  (\ref{e47}) and Table~\ref{tab:1}).  It is
thus tightly constrained already by the empirical lower limit on
$\Gamma^{-1}(\overline{\nu}K^+)$ of $1.9\times 10^{33}$ years to
the point of being disfavored.  For the case of ESSM embedded in
SO(10), the standard $d=5$ operators are suppressed compared to
the case of MSSM; as a result, by themselves they can naturally
lead to lifetimes in the range of $(1-10)\times 10^{33}$ years,
for nearly central values of the parameters pertaining to the
SUSY-spectrum and the matrix element (see Eq.  (\ref{e50-2}) and
Table~\ref{tab:1}).  Including the contribution of the new $d=5$
operators, and allowing for a wide variation of the parameters
mentioned above, one finds that the range of $(10^{33}-2\times10^{34})$
years for proton lifetime is not only very plausible but it also
provides a rather conservative upper limit, for the case of ESSM
embedded in either SO(10) or G(224) (see Section
\ref{Contribution} and Table~\ref{tab:2}). Thus our study provides
a clear reason to expect that the discovery of proton decay should
be imminent for the case of ESSM, and even more so for that of
MSSM.  The implication of this prediction for a next-generation
detector is emphasized in the next section.

\section{\large Concluding Remarks}
\label{Conclusions}

The preceding sections show that, but for two missing
pieces---supersymmetry and proton decay---the evidence in support
of grand unification is now strong.  It includes:  (i) the
observed family-structure, (ii) quantization of electric charge,
(iii) the meeting of the gauge couplings, (iv)
neutrino-oscillations as observed at SuperK, (v) the intricate
pattern of the masses and mixings of all fermions, including the
neutrinos, and (vi) the need for B--L as a generator, to implement
baryogenesis.  Taken together, these not only favor grand
unification but in fact select out a particular route to such
unification, based on the ideas of supersymmetry, SU(4)-color and
left-right symmetry.  Thus they point to the relevance of an
effective string-unified G(224) or SO(10)-symmetry in four
dimensions, as discussed in Sections  3 and 4.

Based on a systematic study of proton decay within the
supersymmetric  SO(10)/G(224)-framework, that (a)
allows for the possibilities of both MSSM and ESSM, and (b) incorporates
the improved values of the matrix element and renormalization effects,
 I have argued that a conservative
upper limit on the proton lifetime is about
(1/3\mbox{-}2)$\times10^{34}$ years.

So, unless the fitting of all the pieces listed above is a mere
coincidence, it is hard to believe that that is the case, discovery
of proton decay should be around the corner.  In particular, as
mentioned in the Introduction, one expects that candidate events
should very likely be observed in the near future already at
SuperK, if its operation is restored.  However, allowing for the
possibility that proton lifetime may well be near the upper limit
stated above, a next-generation detector providing a net
gain in sensitivity by a factor five to ten, compared to SuperK,
would be needed to produce real events and distinguish them
unambiguously from the background.  Such an improved detector
would of course be essential to study the branching ratios of
certain crucial though (possibly) sub-dominant decay modes such as
the $\mu^{+}K^{0}$ and $e^+\pi^0$ as mentioned in Section
\ref{TheCharged}.

The reason for pleading for such improved searches is that proton
decay would provide us with a wealth of knowledge about physics at
truly short distances ($<10^{-30}$ cm), which cannot be gained by
any other means. Specifically, the observation of proton decay, at
a rate suggested above, with $\overline{\nu}K^{+}$ mode being
dominant, would not only reveal the underlying unity of quarks and
leptons but also the relevance of supersymmetry.  It would also
confirm a unification of the fundamental forces at a scale of
order $2\times10^{16}$ GeV. Furthermore, prominence of the
$\mu^{+}K^{0}$ mode, if seen, would have even deeper significance,
in that in addition to supporting the three features mentioned
above, it would also reveal the link between neutrino masses and
proton decay, as discussed in Section \ref{Expectations}.  {\it In
this sense, the role of proton decay in probing into physics at
the most fundamental level is unique}. In view of how valuable
such a probe would be and the fact that the predicted upper limit
on the proton lifetime is at most a factor of three to ten higher
than the empirical lower limit, the argument in favor of building
an improved detector seems compelling.

To conclude, the discovery of proton decay would undoubtedly constitute
a landmark in the history of physics.  It would provide the last,
missing piece of gauge unification and would shed light on how such a
unification may be extended to include gravity in the context of a
deeper theory.

\vskip1.5em

{\bf Acknowledgments:} I would like to thank Kaladi S. Babu and
Frank Wilczek for a most enjoyable collaboration on topics covered
in these lectures. Discussions with Kaladi S. Babu in updating the
results of the previous study have been most helpful.
Correspondence with Michael Dine and Edward Witten on issues
pertaining to the questions of unification and proton decay have
been beneficial.  I am grateful to K. Turznyski for communicating
the details of his results on renormalization effects. I would
like to thank Antonio Masiero, Goran Senjanovic and Alexie Smirnov
for inviting me to lecture at the summer school at the ICTP,
Trieste and also for their kind hospitality. The research
presented here is supported in part by DOE grant no.
DE-FG02-96ER-41015 and by the Department of Energy
    under contract number DE--AC03--76SF00515.

\appendix

\section*{\large APPENDIX: A Natural Doublet-Triplet Splitting Mechanism in
SO(10)}

\setcounter{equation}{0}
\renewcommand{\theequation}{A\arabic{equation}}

In supersymmetric SO(10), a  natural doublet--triplet splitting
can be achieved by coupling the adjoint  Higgs $\mathbf{45_{H}}$
to a $\mathbf{10_{H}}$ and a $\mathbf{10'_{H}}$, with
$\mathbf{45_{H}}$ acquiring a unification--scale VEV in the B--L
direction \cite{DimWilczek,ref:cc}:
$\langle\mathbf{45_{H}}\rangle=(a,a,a,0,0)\times\tau_{2}$ with
$a\sim{M}_{U}$.  As discussed in Section \ref{Understanding}, to
generate  CKM mixing for fermions we require $(\mathbf{16_{H}})_d$
to acquire a VEV of the
 electroweak scale. To ensure accurate gauge coupling unification,
the effective low energy theory should not contain split
multiplets beyond those of MSSM. Thus the MSSM Higgs doublets must
be linear combinations of the SU(2$)_{L}$ doublets in
$\mathbf{10_{H}}$  and $\mathbf{16_{H}}$. A simple set of
superpotential terms that ensures this  and incorporates
doublet-triplet splitting is \cite{BabuWilczekPati}:
\begin{equation}
W_{H}\,=\,\lambda\,\mathbf{10_{H}\,45_{H}\,10'_{H}}\,+\,M_{10}\,
\mathbf{10'_{H}}^{2}\,+\,\lambda'\,\overline\mathbf{16}_{H}\,
\overline\mathbf{16}_{H}\,
\mathbf{10}_{H}\,+\,M_{16}\,\mathbf{16}_{H}\overline{\mathbf{16}}_{H}\,.
\label{a1}
\end{equation}
A complete superpotential for $\mathbf{45_{H}}$,
$\mathbf{16_{H}}$, $\mathbf{\overline{16}_{H}}$,
$\mathbf{10}_{H}$, $\mathbf{10}'_{H}$ and possibly other fields,
which  ensure that (a) $\mathbf{45_{H}}$, $\mathbf{16_{H}}$  and
$\mathbf{\overline{16}_{H}}$ acquire unification scale VEVs with
$\langle\mathbf{45_{H}}\rangle$ being  along the $(B$-$L)$
direction; (b) that exactly two Higgs doublets $(H_{u},H_{d})$
remain light, with $H_{d}$ being a linear combination  of
$(\mathbf{10_{H}})_{d}$ and $(\mathbf{16_{H}})_{d}$; and (c) there
are  no unwanted pseudoGoldstone bosons, can be constructed. With
$\langle\mathbf{45_{H}}\rangle$ in the B--L direction, it does not
contribute to the Higgs doublet mass matrix, so one pair of Higgs
doublet remains light, while all triplets acquire unification
scale masses.  The light MSSM Higgs doublets are
\cite{BabuWilczekPati}
\begin{equation}
H_{u}\,=\,\mathbf{10}_{u}\,,\,\,\,\,H_{d}\,=\,\cos\gamma\,
\mathbf{10}_{d}\,+\,\sin\gamma\,\mathbf{16}_{d}\,, \label{a2}
\end{equation}
with
$\tan\gamma\equiv\lambda'\langle{\bf\overline{16}_{H}}\rangle/M_{16}$.
Consequently, $\langle{\bf10}\rangle_{d}=(\cos\gamma)\,v_{d}$,
$\langle{\bf16}_{d}\rangle=(\sin\gamma)\,v_{d}$, with $\langle
H_{d}\rangle=v_{d}$ and $\langle\mathbf{16}_{d}\rangle$ and
$\langle\mathbf{10}_{d}\rangle$ denoting the electroweak VEVs of
those multiplets. Note that $H_{u}$ is purely in $\mathbf{10}_H$
and that $\left \langle \mathbf{10}_d \right \rangle^2 + \left
\langle \mathbf{16}_d \right \rangle^2 = v_d^2$. This mechanism of
doublet-triplet (DT) splitting is the simplest for the minimal
Higgs systems. It has the advantage that it meets the requirements
of both D-T splitting and CKM-mixing.  In turn, it has three
special consequences:

(i) It modifies the familiar SO(10)-relation $\tan \beta \equiv
v_u/v_d = m_t/m_b \approx 60$ to \footnote{ It is worth noting
that the simple relationship between $\cos\gamma$ and
$\tan\beta$---i.e. $\cos\gamma\approx\tan\beta/(m_t/m_b)$---would
be modified if the superpotential contains an additional term like
$\lambda''\mathbf{16}_H\cdot \mathbf{16}_H\cdot \mathbf{10}_H'$,
which would induce a mixing between the doublets in
$\mathbf{10}'_d$, $\mathbf{16}_d$ and $\mathbf{10} _d$. That in
turn will mean that the upper limit on $M_{\rm eff}\cos\gamma$
following from considerations of threshold corrections (see below)
will not be strictly proportional to $\tan\beta$.  I thank Kaladi
Babu for making this observation.}:
\begin{equation} \tan \beta/\cos \gamma \approx
m_t/m_b \approx 60 \ .
\end{equation}
As a result, even low to
moderate values of $\tan \beta \approx 3$ to 10 (say) are
perfectly allowed in SO(10) (corresponding to $\cos \gamma \approx
1/20$ to $1/6$).

(ii) The most important consequence of the DT-splitting mechanism
outlined above is this:  In contrast to SU(5), for which the
strengths of the standard $d=5$ operators are proportional to
$(M_{H_c})^{-1}$ (where $M_{H_C}\sim few \times 10^{16}$ GeV (see
Eq.  (\ref{e43})), for the SO(10)-model, they become proportional
to $M_{\rm eff}^{-1}$, where $M_{\rm eff} =(\lambda a)^2/M_{10'}
\sim M_X^2/M_{10'}$.  As noted in Ref. \cite{BabuWilczekPati},
$M_{10'}$ can be naturally smaller (due to flavor symmetries) than
$M_X$ and thus $M_{\rm eff}$ correspondingly larger than $M_{X}$
by even one to three orders of magnitude.  Now the proton decay
amplitudes for SO(10) in fact possess an intrinsic enhancement
compared to those for SU(5), owing primarily due to differences in
their Yukawa couplings for the up sector (see Appendix C in Ref.
\cite{BabuWilczekPati}).  As a result, these larger values of
$M_{\rm eff}\sim(10^{18}-10^{19})$ GeV are in fact needed for the
SO(10)-model to be compatible with the observed limit on the
proton lifetime.  At the same time, being bounded above by
considerations of threshold effects (see below), they allow
optimism as regards future observation of proton decay.

(iii) $M_{\rm eff}$ gets bounded above by considerations of coupling
unification and GUT-scale threshold effects as follows.  Let us recall
that in the absence of unification-scale threshold and Planck-scale
effects, the MSSM value of $\alpha_3(m_Z)$ in the $\overline{\mbox{MS}}$
scheme, obtained by assuming gauge coupling unification, is given by
$\alpha_3(m_Z)_{\mbox{\scriptsize{MSSM}}}^\circ = 0.125 - 0.13$
\cite{Langacker}.  This is about 5 to 8\% {\it higher} than the observed
value:  $\alpha_3(m_Z)=0.118\pm 0.003$ \cite{ParticleDataGroup}.  Now,
assuming coupling unification, the net (observed) value of $\alpha_3$,
for the case of MSSM embedded in SU(5) or SO(10), is given by:
\begin{equation}\label{a4} \alpha_3(m_Z)_{\mbox{\scriptsize{net}}}=
\alpha_3(m_Z)_{\mbox{\scriptsize{MSSM}}}^\circ +
\Delta\alpha_3(m_Z)_{\mbox{\scriptsize{DT}}}^{\mbox{\scriptsize{MSSM}}}
+ \Delta_3' \end{equation} where
$\Delta\alpha_3(m_Z)_{\mbox{\scriptsize{DT}}}$ and $\Delta_3'$
represent GUT-scale threshold corrections respectively due to
doublet-triplet splitting and the splittings in the other
multiplets (like the gauge and the Higgs multiplets), all of which
are evaluated at $m_Z$.  Now, owing to mixing between
$\mathbf{10}_d$ and $\mathbf{16}_d$ [see Eq.  (\ref{a2})], one
finds that $\Delta\alpha_3(m_Z)_{\mbox{\scriptsize{DT}}}$ is given
by $[\alpha_3(m_Z)^2/2\pi](9/7)\ln (M_{\mbox{\scriptsize{\rm
eff}}} \cos\gamma/M_X)$ \cite{BabuWilczekPati}.

As mentioned above, constraint from proton lifetime sets a lower
limit on $M_{\mbox{\scriptsize{\rm eff}}}$ given by
$M_{\mbox{\scriptsize{\rm eff}}}> (1-6)\times 10^{18}$GeV.  Thus,
even for small $\tan\beta\approx 2$ (i.e.  $\cos\gamma \approx
\tan(\beta/60) \approx 1/30$),
$\Delta\alpha_3(m_Z)_{\mbox{\scriptsize{DT}}}$ is positive; and it
increases logarithmically with $M_{\mbox{\scriptsize{\rm eff}}}$.
Since $\alpha_3(m_Z)_{\mbox{\scriptsize{MSSM}}}^\circ$ is higher
than $\alpha_3(m_Z)_{\mbox{\scriptsize{obs}}}$, and as we saw,
$\Delta\alpha_3(m_Z)_{\mbox{\scriptsize{DT}}}$ is positive, it
follows that the corrections due to {\it other} multiplets denoted
by $\delta_3'=\Delta_3'/\alpha_3(m_Z)$ should be appropriately
negative so that $\alpha_3(m_Z)_{\mbox{\scriptsize{net}}}$ would
agree with the observed value.

In order that coupling unification may be regarded as a natural
prediction of SUSY unification, as opposed to being a mere
coincidence, it is important that the magnitude of the net other
threshold corrections, denoted by $\delta_3'$, be negative but not
any more than about 8 to 10\% in magnitude (i.e.  $-\delta_3' \leq
(8-10)\%$). It was shown in Ref.  \cite{BabuWilczekPati} that the
contributions from the gauge and the minimal set of Higgs
multiplets (i.e.  $\mathbf{45}_H, \mathbf{16} _H,
\overline{\mathbf{16}}_H$ and $\mathbf{10}_H$) leads to threshold
correction, denoted by $\delta_3'$, which has in fact a negative
sign and quite naturally a magnitude of 4 to 8\%, as needed to
account for the observed coupling unification.  The correction to
$\alpha_3(m_Z)$ due to Planck scale physics through the effective
operator $F_{\mu\nu}F^{\mu\nu}\mathbf{45}_H/M$ does not alter the
estimate of $\delta_3'$ because it vanishes due to antisymmetry in
the SO(10)- contraction.

Imposing that $\delta_3'$ (evaluated at $m_Z$)be negative and not
any more than about 10-11\% in magnitude in turn provides a
restriction on how big the correction due to doublet-triplet
splitting---i.e.
$\Delta\alpha_3(m_Z)_{\mbox{\scriptsize{D$\bar{T}$}}}$---can be.
That in turn sets an upper limit on $M_{\rm eff}\cos\gamma$, and
thereby on $M_{\rm eff}$ for a given $\tan\beta$.  For instance,
for MSSM, with $\tan\beta=(2,3,8)$, one obtains (see Ref.
\cite{BabuWilczekPati}): $M_{\rm eff}\le(4,2.66,1)\times 10^{18}$
GeV.  Thus, conservatively, taking $\tan\beta\ge 3$, one obtains:
\begin{equation} M_{\rm eff} \lsim 2.7\times 10^{18} \mbox{GeV
(MSSM)} \quad (\tan\beta\geq 3) \ .
\end{equation}

\subsection*{Limit on \boldmath $M_{\rm eff}$ \unboldmath
For The case of ESSM }

Next consider the restriction on $M_{\rm eff}$ that would arise
for the case of the extended supersymmetric standard model (ESSM),
which introduces an extra pair of vector-like families
($16+\bar{16})$ of SO(10)) at the TeV scale \cite{BabuJi}(see also
footnote 16).  In this case, $\alpha_{\mbox{\scriptsize{unif}}}$
is raised to 0.25 to 0.3, compared to 0.04 in MSSM.  Owing to
increased two-loop effects the scale of unification M$_X$ is
raised to $(1/2-2)\times 10^{17}$ GeV, while
$\alpha_3(m_Z)_{\mbox{\scriptsize{ESSM}}}^\circ$ is lowered to
about 0.112-0.118 \cite{BabuJi,KoldaRussell}.

With raised M$_X$, the product $M_{\rm eff}\cos\gamma \approx
M_{\rm eff}(\tan\beta)/60$ can be higher by almost a factor of
five compared to that for MSSM, without altering
$\Delta\alpha_3(m_Z)_{\mbox{\scriptsize{DT}}}$.  Furthermore,
since $\alpha_3(m_Z)_{\mbox{\scriptsize{MSSM}}}^\circ$ is
typically lower than the observed value of $\alpha_3(m_Z)$
(contrast this with the case of ESSM), for ESSM, $M_{\rm eff}$ can
be higher than that for MSSM by as much as a factor of 2 to 3,
without requiring an enhancement of $\delta_3'$. The net result is
that for ESSM embedded in SO(10), $\tan\beta$ can span a wide
range from 3 to even 30 (say) and simultaneously the upper limit
on $M_{\rm eff}$ can vary over the range (60 to 6)$\times 10^{18}
GeV$, satisfying \begin{equation}\label{MeffEq} M_{\rm eff}\lsim
(6\times 10^{18} \mbox{GeV})(30/\tan\beta) \mbox{\,\,(ESSM)},
\end{equation} with the unification-scale threshold corrections
from ``other'' sources denoted by
$\delta_3'=\Delta_3'/\alpha_3(m_Z)$ being negative, but no more
than about 5\% in magnitude.  As noted above, such values of
$\delta_3'$ emerge quite naturally for the minimal Higgs system.
Thus, one important consequence of ESSM is that by allowing for
larger values of $M_{\rm eff}$ (compared to MSSM), without
entailing larger values of $\delta_3'$, it can be perfectly
compatible with the limit on proton lifetime for almost {\it
central values} of the parameters pertaining to the SUSY spectrum
and the relevant matrix elements (see Eq. (40)). Further, larger
values of $\tan\beta$ (10 to 30, say) can be compatible with
proton lifetime only for the case of ESSM, but not for MSSM.
These features are discussed in the text, and also exhibited in
Table~\ref{tab:2}.

\newpage

\begin{table}[htb]
{\bf\caption[*]{\bf \large Values of proton lifetime
    \boldmath
    $\left(\Gamma^{-1}(p\rightarrow\bar{\nu}K^+)\right)$\unboldmath
   for a wide range of parameters.\label{tab:2}}}\medskip
\begin{tabular}{|c|c|c||c|c||c|}
\hline \rule[-3mm]{0mm}{8mm} Parameters  &
\multicolumn{2}{|c||}{MSSM $\to$ SO(10)}
        & \multicolumn{2}{|c||}{ESSM $\to$ SO(10)}
        &                           \\
(spectrum/Matrix  &  \multicolumn{2}{|c||}{{\bf Std. d=5}}
            &  \multicolumn{2}{|c||}{{\bf Std. d=5}}
        &  \raisebox{1.5ex}[0pt] {$\left\{\begin{tabular}{c}
    \vspace*{-1ex} MSSM \\ or \vspace*{-0.6ex} \\ ESSM \end{tabular}
        \right\}\to$ G(224)/SO(10)} \\
element)     & \multicolumn{2}{|c||}
            {Intermed. $\epsilon'$ \& phase$^\dagger$}
         & \multicolumn{2}{|c||}
            {Intermed. $\epsilon'$ \& phase$^\dagger$}
         & {\bf New d=5}$^{\dagger\dagger}$     \\ \cline{2-6}
             & $\tan\beta$=$3$ & $\tan\beta$=$10$
         & $\tan\beta$=$5$ & $\tan\beta$=$10$
             & Independent of $\tan\beta$       \\ \hline
Nearly ``central'' & $0.2\times10^{32}$  & 1.6$\times 10^{30}$
         & 0.25$\times 10^{34}$  & 0.7$\times10^{33}$
         & $0.50\times 10^{33}$                 \\
\{C\}=2  & yrs & yrs & yrs$^{{*}}$
         & yrs & yrs$^{\dagger\dagger}$ \\ \hline
Intermediate     & 0.7$\times 10^{32}$  & 0.6$\times 10^{31}$
         & 1$\times 10^{34}$  & 2.8$\times 10^{33}$
         & 2$\times 10^{33}$          \\
\{C\}=8  & yrs & yrs & yrs$^{{*}}$
         & yrs & yrs$^{\dagger\dagger}$ \\ \hline
Nearly Extreme   & 0.3$\times 10^{33}$  & 2.6$\times 10^{31}$
         & 4$\times 10^{34}$  & 1.1$\times 10^{34}$
         & 8$\times 10^{33}$              \\
\{C\}=32 & yrs & yrs & yrs$^{{*}}$
         & yrs & yrs$^{\dagger\dagger}$ \\ \hline
\end{tabular}
\\[2ex]
 ${*}${\bf In this case, lifetime is given by the last
column.}
\end{table}

$\bullet$ Since we are interested in exhibiting expected proton
lifetime near the upper end, we are not showing entries in
Table~\ref{tab:2} corresponding to values of the parameters for
the SUSY spectrum and the matrix element [see Eq. (\ref{e39}), for
which the curly bracket \{C\} appearing in Eqs. (\ref{e46}),
(\ref{e49-2}), (\ref{e52})] would be less than one (see however
Table~\ref{tab:1}). In this context, we have chosen here ``nearly
central'', ``intermediate'' and ``nearly extreme'' values of the
parameters such that the said curly bracket is given by 2, 8 and
32 respectively, instead of its extreme upper-end value of 64. For
instance, the curly bracket would be 2 if $\beta_H=(0.0117)$
GeV$^3$, $m_{\tilde{q}}\approx 1.2$ TeV and
$m_{\tilde{W}}/m_{\tilde{q}}\approx (1/7.2)$, while it would be 8
if $\beta_H=0.010$ GeV$^3$, $m_{\tilde{q}}\approx 1.44$ TeV and
$m_{\tilde{W}}/m_{\tilde{q}}\approx 1/10$; and it would be 32 if,
for example, $\beta_H=0.007$ GeV$^3$, $m_{\tilde{q}}\approx
\sqrt{2}(1.2$ TeV) and $m_{\tilde{W}}/m_{\tilde{q}}\approx 1/12$.

$\mathbf{\dagger}$ All the entries for the standard $d=5$
operators correspond to taking an intermediate value of
$\epsilon'\approx (1$ to $1.4)\times 10^{-4}$ (as opposed to the
extreme values of $2\times 10^{-4}$ and zero for cases I and II,
see Eq. (\ref{e33})) and an intermediate phase-dependent factor
such that the uncertainty factor in the square bracket appearing
in Eqs. (\ref{e46}) and (\ref{e49-2}) is given by 5, instead of
its extreme values of $2\times 4=8$ and $2.5\times 4=10$,
respectively.

$\mathbf{\dagger\dagger}$ For the new operators, the factor
[8-1/64] appearing in Eq. (\ref{e52}) is taken to be 6, and
$K^{-2}$, defined in Section \ref{Preliminaries}, is taken to be
25, which are quite plausible, in so far as we wish to obtain
reasonable values for proton lifetime at the upper end.

$\bullet$ The standard $d=5$ operators for both MSSM and ESSM are
evaluated by taking the upper limit on $M_{\rm eff}$ (defined in
the text) that is allowed by the requirement of natural coupling
unification. This requirement restricts threshold corrections and
thereby sets an upper limit on $M_{\rm eff}$, for a given
$\tan\beta$ (see Section \ref{Expectations} and Appendix).

${\mathbf \ast}$ For all cases, the standard and the new $d=5$
operators must be combined to obtain the net amplitude.  For the
three cases of ESSM marked with an asterisk, and other similar
cases which arise for low $\tan\beta\approx 3$ to 6 (say), the
standard $d=5$ operators by themselves would lead to proton
lifetimes typically exceeding $(0.25\mbox{-}4)\times10^{34}$
years.  For these cases, however, the contribution from the new
$d=5$ operators would dominate, which quite naturally lead to
lifetimes in the range of $(10^{33}-10^{34})$ years (see last
column).

$\bullet$ As shown above, the case of MSSM embedded in SO(10) is tightly
constrained to the point of being disfavored by present empirical lower
limit on proton lifetime Eq.  (\ref{e41}) [see discussion following Eq.
(\ref{e47})].

$\bullet$ Including contributions from the standard and the new
operators, the case of ESSM, embedded in either G(224) or SO(10), is,
however, fully consistent with present limits on proton lifetime for a
wide range of parameters; at the same time it provides optimism that
proton decay will be discovered in the near future, with a lifetime
$\leq 10^{34}$ years.

$\bullet$ The lower limits on proton lifetime are not exhibited.  In the
presence of the new operators, these can typically be as low as about
$10^{29}$ years (even for the case of ESSM embedded in SO(10)).  Such
limits and even higher are of course long excluded by experiments.

\end{document}